\documentclass[epj]{svjour-pre}
\usepackage{amsmath}
\usepackage{amssymb}
\usepackage[nosort]{cite}
\usepackage[dvips]{graphicx}
\usepackage{epsfig}
\usepackage{rotating}
\usepackage{url}

\newcommand{\andEq}[1]{(\ref{#1})}
\newcommand{\Eq}[1]{Eq.~(\ref{#1})}
\newcommand{\Eqs}[1]{Eqs.~(\ref{#1})}
\newcommand{\Equation}[1]{Equation~(\ref{#1})}
\newcommand{\Fig}[1]{Fig.~\ref{#1}}
\newcommand{\Figure}[1]{Figure~\ref{#1}}
\newcommand{\Ref}[1]{Ref.~\citen{#1}}
\newcommand{\Refs}[1]{Refs.~\citen{#1}}
\newcommand{\Sec}[1]{Sect.~\ref{#1}}
\newcommand{\Secs}[1]{Sects.~\ref{#1}}
\newcommand{\Tab}[1]{Table~\ref{#1}}
\newcommand{\Tabs}[1]{Tables~\ref{#1}}
\newcommand{\Table}[1]{Table~\ref{#1}}

\newcommand{\BR}[1]{\ensuremath{{\rm BR}(#1)}}
\newcommand{\dEM}[1]{\ensuremath{\delta_{\rm EM}^{#1}}}
\newcommand{\dSU}[1]{\ensuremath{\delta_{\rm SU(2)}^{#1}}}
\newcommand{\NF}[1]{\ensuremath{N_F=#1}}
\newcommand{\order}[1]{\ensuremath{\mathcal{O}(#1)}}
\newcommand{\SN}[2]{\ensuremath{#1\times10^{#2}}}
\newcommand{\SU}[1]{\ensuremath{\rm SU(#1)}}
\newcommand{\U}[1]{\ensuremath{\rm U(#1)}}

\newcommand{\fp}{\ensuremath{f_+(0)}}
\newcommand{\FK}{\ensuremath{f_K}}
\newcommand{\FKpi}{\ensuremath{f_K/f_\pi}}
\newcommand{\Fpi}{\ensuremath{f_\pi}}
\renewcommand{\Im}{\ensuremath{\mathrm{Im}}}
\newcommand{\lnC}{\ensuremath{\ln C}}
\newcommand{\lp}{\ensuremath{\lambda_+}}
\newcommand{\Lp}{\ensuremath{\Lambda_+}}
\newcommand{\lz}{\ensuremath{\lambda_0}}

\newcommand{\stat}{\ensuremath{_\mathrm{stat}}}
\newcommand{\syst}{\ensuremath{_\mathrm{syst}}}
\newcommand{\Vub}{\ensuremath{|V_{ub}|}}
\newcommand{\Vud}{\ensuremath{|V_{ud}|}}
\newcommand{\Vus}{\ensuremath{|V_{us}|}}
\newcommand{\Vusd}{\ensuremath{|V_{us}/V_{ud}|}}
\newcommand{\Vusf}{\ensuremath{|V_{us}|f_+(0)}}

\mathchardef\mhyp="2D

\begin{document}

%%%%%%%%%%%%%%%%%%%%%%%%%%%%%%%%%%%%%%%%%%%%%%%%%%%%%%%%%%%%%%%%%%%%%%%%%%%%
%% PREPRINT NUMBERS (to be added in Journal-Refs box for arXiv submission %%
%% ICCUB-10-021, UB-ECM-PF-10-009, IFIC/10-12                             %%
%%%%%%%%%%%%%%%%%%%%%%%%%%%%%%%%%%%%%%%%%%%%%%%%%%%%%%%%%%%%%%%%%%%%%%%%%%%%

\def\preprintnumbers
{ICCUB-10-021, UB-ECM-PF-10-009, IFIC/10-12}

\title{\mathversion{bold}An evaluation of \Vus\ and precise 
tests of the Standard Model from world data on leptonic and
semileptonic kaon decays}

\titlerunning{Evaluation of \Vus\ and Standard Model tests from kaon data}

\author{
M.~Antonelli\inst{2},
V.~Cirigliano\inst{3},
G.~Isidori\inst{2},
F.~Mescia\inst{1},
M.~Moulson\inst{2},
H. Neufeld\inst{8},
E.~Passemar\inst{7},
M.~Palutan\inst{2},
B.~Sciascia\inst{2},
M.~Sozzi\inst{5},
R.~Wanke\inst{4},
O.P.~Yushchenko\inst{6},
for the FlaviaNet Working Group on Kaon Decays\inst{9}
}

\institute{
Dep.\ ECM and ICC, Universitat de Barcelona, 08028 Barcelona, Spain \and
Laboratori Nazionali di Frascati dell'INFN, 00044 Frascati RM, Italy \and
Theoretical Division, Los Alamos National Laboratory, Los Alamos, NM, 87545, 
USA \and
Institut f\"ur Physik, Universit\"at Mainz, 55099 Mainz, Germany \and
Dipartimento di Fisica, Universit\`a di Pisa e Sezione dell'INFN di Pisa, 56100 Pisa, Italy \and
Institute for High Energy Physics, 142284 Protvino, Russia \and
Departament de F\'isica Te\`orica, IFIC, Universitat de Val\`encia - CSIC,
46071 Val\`encia, Spain \and
Fakult\"at f\"ur Physik, Universit\"at Wien, 1090 Wien, Austria \and
\texttt{http://www.lnf.infn.it/wg/vus}
}

\authorrunning{M. Antonelli et al.}

\mail{Matthew.Moulson@lnf.infn.it}

\date{Received: date / Revised version: date}

\abstract{We present a global analysis of leptonic and semileptonic kaon 
decay data, including all recent results published by the BNL-E865, KLOE, 
KTeV, ISTRA+ and NA48 experiments. This analysis, in conjunction with 
precise lattice calculations of the hadronic matrix elements now available, 
leads to a very precise determination of \Vus\ and allows us to perform 
several stringent tests of the Standard Model.
}

\maketitle

\section{Introduction}
\label{sec:intro}

Within the Standard Model (SM), leptonic and semileptonic kaon decays
can be used to obtain the most accurate determination of the magnitude of
the element $V_{us}$ of the Cabibbo-Kobayashi-Maskawa (CKM) 
matrix \cite{Cabibbo:1963yz,Kobayashi:1973fv}.
A detailed analysis of these processes potentially also provides
stringent constraints on new physics scenarios:
while within the SM, all $d^i \to  u^j \ell \nu$ transitions 
are ruled by the same CKM coupling $V_{ji}$ (satisfying
the unitarity condition $\sum_k |V_{jk}|^2 =1$), and 
$G_F$ is the same coupling that governs muon decay, 
this is not necessarily true beyond the SM. 
New bounds on violations of CKM unitarity and lepton universality 
and deviations from the $V-A$ structure translate into significant 
constraints on various new-physics scenarios. Alternately, such tests may
eventually turn up evidence of new physics. 

In the case of leptonic and semileptonic kaon decays, these tests 
are particularly significant given (i) the large amount of
data recently collected by several experiments,
(ii) the substantial progress recently made in evaluating 
the corresponding hadronic matrix elements from lattice QCD, and
(iii) the precise analytic calculations of radiative 
corrections and isospin-breaking effects recently performed
within chiral perturbation theory (ChPT), the low-energy effective 
theory of QCD.
This progress on both the experimental and the theoretical 
sides allows for unique tests of the SM that probe 
very high energy scales.

An illustration of the importance
of semileptonic kaon decays in testing the SM 
is provided by the unitarity relation 
\begin{equation}
\Vud^2 + \Vus^2 + \Vub^2 = 1 + \Delta_{\rm CKM}.
\label{eq:unitarity}
\end{equation}
Here the $V_{ji}$ are the CKM elements as determined from 
the various $d^i \to  u^j$ processes, where the value of $G_F$ is determined 
from the muon life time: 
$G_\mu = 1.166371 (6) \times 10^{-5} {\rm GeV}^{-2}$ \cite{Chitwood:2007pa}.
$\Delta_{\rm CKM}$ parameterizes possible 
deviations from the SM induced by dimension-six operators, 
contributing either to muon decay or to $d^i \to  u^j$ 
transitions. As we will show in the following, 
the present accuracy on \Vus\ allows us  
to set bounds on $\Delta_{\rm CKM}$ around $0.1\%$,
which translate into bounds on the effective scale of new physics
on the order of 10~TeV.

A detailed analysis of precise tests of the SM 
with leptonic and semileptonic kaon decays has already 
been presented in \Ref{Antonelli:2008jg}. 
However, the significant progress on both the experimental and theoretical 
sides has motivated us to perform an updated analysis with three major areas
of emphasis: (i) the determination of \Vus\ from experimental data, 
with and without imposing CKM unitarity; (ii) the comparison between 
the values of \Vus\ obtained from data on $K\to\pi\ell\nu$ ($K_{\ell3}$) 
and $K\to\mu\nu$ ($K_{\mu2}$) decays and the corresponding 
constraints on deviations from the $V-A$ structure of the charged current;
(iii) tests of lepton universality in $K_{\ell3}$ decays.

\begin{sloppypar}
To carry out this analysis, values are needed for the hadronic constants
\FKpi\ and \fp, as discussed in \Secs{sec:FKpi} and~\ref{sec:fp}. These
values are obtained using lattice QCD, and various determinations have been 
performed. The lattice QCD community, as represented by the FlaviaNet 
Lattice Averaging Group (FLAG) \cite{Colangelo:2009,Lubicz:2010nx,FLAG1}, 
is progressing towards convergence on a set of reference values, 
but in the meantime, for the 
purposes of this work, we are led to propose our own.
The criteria we have applied in averaging lattice QCD results are motivated, 
but do not represent the only set of possible choices. Our adoption of 
these values is intended to illustrate what precision can be obtained 
in testing the SM, given current experimental and 
theoretical results. In particular, wherever possible we quote results
for quantities such as \Vusf\ or $\Vusd\times\FKpi$, which are independent
of lattice inputs and ready for use as new lattice results become available.
\end{sloppypar} 

This paper is organized as follows. The phenomenological
framework needed to describe $K_{\ell3}$ and $K_{\mu2}$ decays 
within and beyond the SM is briefly reviewed in \Sec{sec:theo}.  
The experimental data is reviewed and combined in \Sec{sec:data}.
The results are presented and interpreted in \Sec{sec:res}.

\section{Phenomenological framework}
\label{sec:theo}

\subsection{\mathversion{bold}$K_{\ell2}$ rates in the Standard Model}
\label{sec:Pl2}

\begin{sloppypar}
Within the SM, the ratio of photon-inclusive 
$K^\pm\to\ell^\pm\nu$ ($K^\pm_{\ell2(\gamma)}$) to $\pi^\pm\to\ell^\pm\nu$ 
($\pi^\pm_{\ell2(\gamma)}$) 
decay rates can be written as \cite{Marciano:1993sh,Cirigliano:2007ga}
\begin{equation}
\frac{\Gamma_{K_{\ell2}}}{\Gamma_{\pi_{\ell2}}} =
\frac{\Vus^2}{\Vud^2}\frac{\FK^2}{\Fpi^2}
\frac{m_K(1 - m_\ell^2/m_K^2)^2}{m_\pi(1 - m_\ell^2/m_\pi^2)^2} 
\left(1 + \dEM{}\right),
\label{eq:Mkl2}
\end{equation}
where \FK\ and \Fpi\ are the kaon and pion decay constants, and
\dEM{} denotes the effect of long-distance electromagnetic
corrections. Short-distance radiative effects are universal and cancel from 
the ratio. For pointlike kaons and pions, the long-distance electromagnetic 
corrections depend only on the particle masses. The dominant uncertainty
on \dEM{} comes from terms depending on the hadronic structure.
Most analyses to date make use of the results of 
\Refs{Decker:1994ea} and~\citen{Finkemeier:1995gi}, which were computed in a 
model of hadronic structure assuming Breit-Wigner form factors for the 
low-lying vector resonances in order to handle the scale matching. 
These results give $\dEM{} = -0.0070(35)$ (see, e.g., \Ref{Marciano:2004uf}).
Using chiral perturbation theory \cite{Knecht:1999ag,Cirigliano:2007ga}
it has been shown that to leading nontrivial order\footnote
{In ChPT, physical amplitudes are systematically
expanded in powers of the external momenta of pseudo-Goldstone 
bosons $(\pi,K,\eta)$ and quark masses. When including electromagnetic 
corrections, the power counting is in $e^{2m}\,(p/4\pi\Fpi)^{2n}$. Powers of
the quark masses count as two powers of the external momenta 
($\order{p^2} = \order{m_q}$).}
$\order{e^2p^2}$, the structure-dependent corrections to \dEM{} 
can be expressed in terms of the electromagnetic pion mass splitting.
With the relative theoretical uncertainty estimated at 25\% 
to account for \order{e^2p^4} effects suppressed by chiral power counting,
one obtains
\begin{equation}
\dEM{} = -0.0070(18).
\end{equation}
\end{sloppypar}

With experimental measurements of the inclusive $K_{\ell2}$ and $\pi_{\ell2}$
decay rates and precise knowledge of the radiative corrections, \Eq{eq:Mkl2} 
can be used to obtain the value of the ratio
\begin{equation}
\left|\frac{V_{us}}{V_{ud}}\right|^2 \frac{f^2_K}{f^2_\pi}.
\end{equation}

\subsubsection{Theoretical determination of \FKpi}
\label{sec:FKpi}

To experimentally constrain the ratio \Vusd, or ultimately, the value of
\Vus\ itself, a precise estimate of the ratio of decay constants
\FKpi\ is needed. The analytic evaluation of this ratio within ChPT at 
$\order{p^4}$ depends on unknown low-energy constants (LECs) and thus 
cannot be predicted with high accuracy. Consequently, precise
evaluations of \FKpi\ are obtained only from lattice QCD.
However, ChPT still provides useful information on this ratio: 
the \SU{3} breaking of \FKpi\ is linear in $m_K^2-m_\pi^2 \propto m_s-m_u$ 
and thus potentially large. For lattice determinations of \FKpi,
this implies that the use of very light pions is essential to obtain 
reliable results.

\begin{sloppypar}
During the last few years, new simulations with \NF{2}, \NF{2+1}, and 
\NF{2+1+1} flavors of dynamical quarks have been performed
by several groups using many different lattice QCD formulations, 
such as staggered (MILC \cite{Bazavov:2009bb}), 
domain-wall (RBC \cite{Allton:2008pn}), 
overlap (JLQCD \cite{JLQCD:2009sk}), and 
Wilson-like fermions (BMW \cite{Durr:2008zz}, CERN-TOV/CLS 
\cite{DelDebbio:2007pz}, PACS-CS \cite{Aoki:2008sm}, and ETMC 
\cite{Baron:2009wt}). 
The fundamental characteristic of these new-generation unquenched studies 
is that recent technical and conceptual
developments \cite{Jansen:2008vs} have allowed pion masses well below
$300$~MeV to be reached with large physical volumes ($L$ up to 4~fm). 
The PACS-CS \cite{Aoki:2008sm} collaboration for example has already
simulated pions as light as $m_\pi=156$~MeV for \NF{2+1} 
(degenerate $u$ and $d$ quarks) and clover (first-principle lattice QCD) 
fermions. The resulting PACS-CS value, $\FKpi=1.189(20)$ \cite{Aoki:2008sm}, 
however, is still plagued by large uncertainties due
to the small simulated volume, $L m_\pi \gtrsim 2.3$, with
corresponding finite-size effects 
$\delta_{\rm {FSE}}=\exp(-L m_\pi) \approx 10\%$.
\end{sloppypar}

\begin{figure*}[t]
\begin{center}
\includegraphics[width=0.63\textwidth]{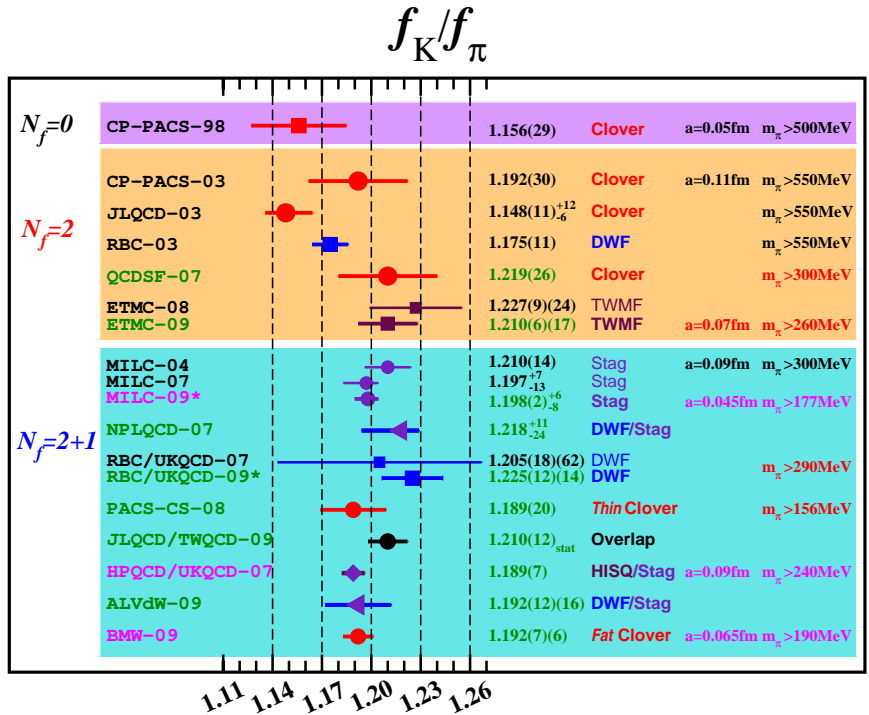}
\caption{\label{fig:fkpi}Summary of lattice determinations of \FKpi\
\cite{Aoki:1999yr,AliKhan:2001tx,Aoki:2002uc,Aoki:2004ht,Gockeler:2006ns,Blossier:2007vv,Blossier:2009bx,Aubin:2004fs,Bernard:2007ps,Bazavov:2009fk,Beane:2006kx,Allton:2008pn,Mawhinney:2009jy,Aoki:2008sm,Noaki:2009sk,Follana:2007uv,Aubin:2008ie,Durr:2010hr}. 
The smallest lattice spacing $a$ (when available), 
lightest value of $m_\pi$ simulated, and type of lattice fermions used 
are listed in each case 
All results for which a value of $a$ is listed have been extrapolated 
by the authors to the contiunuum limit. Where two types of lattice 
fermions are listed (e.g. ``DWF/Stag''), valence and sea quarks are 
discretized differently. Unpublished results are indicated by asterisks.}
\end{center}
\end{figure*}

The present status of lattice results for \FKpi\
\cite{Aoki:1999yr,AliKhan:2001tx,Aoki:2002uc,Aoki:2004ht,Gockeler:2006ns,Blossier:2007vv,Blossier:2009bx,Aubin:2004fs,Bernard:2007ps,Bazavov:2009fk,Beane:2006kx,Allton:2008pn,Mawhinney:2009jy,Aoki:2008sm,Noaki:2009sk,Follana:2007uv,Aubin:2008ie,Durr:2010hr}
is summarized in \Fig{fig:fkpi}. The lightest pion mass simulated is
listed for each determination, together with the smallest lattice spacing 
used, whenever available.
The agreement between the different results is remarkable. 
The present overall accuracy is about 1\%. 

Among the results for \FKpi\ in \Fig{fig:fkpi}, particularly noteworthy 
are the \NF{2+1} studies from BMW \cite{Durr:2010hr}, MILC~'09
\cite{Bazavov:2009fk} and HPQCD/UKQCD \cite{Follana:2007uv}, 
for which detailed evaluation of the systematic errors 
(chiral, continuum and infinite-volume limit) 
have been completed.

The recent study of \FKpi\ from BMW \cite{Durr:2010hr}
involves simulations of $N_F=2+1$ clover fermions at several volumes
with $L m_\pi\gtrsim 4$, whereas the continuum limit has been estimated by 
using three different lattice spacings 
($a \approx 0.065$, 0.085, and 0.125~fm).
The BMW value \cite{Durr:2010hr} is     
\begin{equation}
\FKpi=1.192(7)\stat(6)\syst.
\label{eq:bmw}
\end{equation}
An interesting observation from this study  
is that the value of \FKpi\ at their lightest pion mass
($m_\pi=190$~MeV at $a=0.085$~fm) is only $2\%$ below  
the extrapolated value quoted in \Eq{eq:bmw}.
This strengthens the statement that the result is a
direct QCD measurement, and not an artificial value 
cured by an ad hoc fit.

The 2009 MILC result \cite{Bazavov:2009fk}, in contrast to the BMW result, was
obtained using staggered (AsqTad) fermions.  
Dynamical AsqTad-staggered ensembles have been generated by MILC since 
2002 \cite{Aubin:2004fs,Bernard:2007ps} and now represent 
an ample data set for the study of lattice systematics featuring 
six different lattice spacings (ranging from $a=0.18$~fm down to $a=0.045$~fm), 
light pions ($m_\pi\gtrsim 177$~MeV), and large volumes ($L m_\pi\gtrsim 4$). 
This MILC activity marked the end of the era of quenched calculations  
and the start of high-precision lattice QCD with light pions.
In their 2009 analysis of \FKpi\ \cite{Bazavov:2009fk}, MILC exploits
the new lattice ensembles with $a=0.06$~fm ($m_\pi\gtrsim 224$~MeV)
and $a=0.045$~fm ($m_\pi\gtrsim 324$~MeV), together with the 
older $a=0.09$~fm ($m_\pi\gtrsim 180$~MeV) data set.
The resulting value of \FKpi\ \cite{Bazavov:2009fk} is 
\begin{equation}
\FKpi = 1.198(2)\stat(^{+6}_{-8})\syst = 1.197(7),
\label{eq:milc09}
\end{equation} 
which is in good agreement with the BMW result (\Eq{eq:bmw}). (We perform 
the symmetrization of the total uncertainties in \Eq{eq:milc09} to facilitate
error propagation.) However, the 
sizable shift with respect to the earlier 2004 MILC result \cite{Aubin:2004fs}, 
which made use of a coarser lattice and heavier pions than were available in 
2009, should be noted.
This is a consequence of the sizable systematic errors affecting \FKpi:
significant shifts in the central values 
can arise when going to lighter pion masses and the continuum limit 
(consider also the ETMC results \cite{Blossier:2009bx,Blossier:2007vv} in the
\NF{2} case). 
For this reason, we consider to be exploratory (although important) 
the studies of \FKpi\ in which essentially a single lattice spacing is used,
including the results shown in \Fig{fig:fkpi} from 
RBC/UKQCD \cite{Mawhinney:2009jy}, JLQCD-TWQCD \cite{JLQCD:2009sk},
NPLQCD \cite{Beane:2006kx}, PACS-CS \cite{Aoki:2008sm}, and \Ref{Aubin:2008ie}.

\begin{sloppypar}
Lastly, for the study from HPQCD/UKQCD \cite{Follana:2007uv}
the MILC $a=0.15$~fm, 0.12~fm, and 0.09~fm ensembles are used;
as in the study from MILC, sea quarks are 
treated using AsqTad staggered fermions. 
However, in contrast to the MILC approach, valence quarks are described 
using the Highly-Improved Staggered Quark (HISQ) formulation. The HISQ QCD
action \cite{Follana:2007uv} is understood to give better lattice 
scaling---indeed, the MILC Collaboration itself is 
generating new ensembles with the HISQ action \cite{Bazavov:2010ru}.  
The \FKpi\ value from HPQCD/UKQCD \cite{Follana:2007uv} is 
\begin{equation}
\label{eq:hpqcd}
\FKpi = 1.189(2)\stat(7)\syst.
\end{equation} 
This result is in good agreement with the BMW and MILC results.
In particular, there is no apparent systematic difference between
the results obtained using staggered and clover fermions.
This seems to suggest that possible issues associated with the use of 
staggered fermions (in particular, the rooting 
issue \cite{Creutz:2008nk,Kronfeld:2007ek,Creutz:2007rk})
are not relevant to the determination of \FKpi, at least at the present
level of accuracy.
\end{sloppypar} 
 
In order to fully exploit the data set in \Fig{fig:fkpi}, we average the
results of the analyses from BMW, MILC~'09, and HPQCD/UKQCD discussed above
(\Eqs{eq:bmw}, \andEq{eq:milc09}, and~\andEq{eq:hpqcd}).
Since these results are consistent, we calculate the average weighted
by the statistical errors on the individual results.
This gives our reference central value and its statistical error.
To obtain the total error on this average, we assume a systematic error of 
0.006, equal to the smallest systematic error quoted among the three inputs.
This is justified on the basis of the agreement between the results.
Adding the statistical and systematic errors in quadrature, we obtain
\begin{equation}
\FKpi=1.193(6),
\label{eq:avg}
\end{equation} 
which is quite consistent with all the results in \Fig{fig:fkpi}, 
including those obtained with staggered fermions and from preliminary studies. 
In the above average, possible correlations between the HPQCD/UKQCD and 
MILC results due to the use of a common ensemble with
$a=0.09$~fm has been neglected. 
However, since the valence quarks are treated differently in these two
studies, and the analyses are completely different, any potential
correlations are diluted.  
The above average is consistent with, but has a smaller total uncertainty 
than, both the average from the most recent Lattice 
conference~\cite{Lubicz:2010nx}, $\FKpi=1.196(10)$ and the preliminary FLAG 
result \cite{FLAG1}, $\FKpi=1.190(10)$. As we note in \Sec{sec:intro},
our use of this average to obtain the results presented in \Secs{sec:univ}
and~\ref{sec:bounds} represents a scientific choice. Our value for 
$\Vusd\times\FKpi$ (\Eq{eq:vusvudres})
may be used with an alternate choice for $\FKpi$ to rederive the
results of \Secs{sec:univ} and~\ref{sec:bounds}, if desired.

Updates from PACS-CS, RBC/UKQCD, and JLQCD, in addition to new results
(for example, an \NF{2+1+1} result from ETMC \cite{Baron:2010bv}),
have already been announced and will soon improve the present situation.

\subsection{\mathversion{bold}$K_{\ell3}$ rates in the Standard Model}
\label{sec:Kl3}

\begin{sloppypar}
The $K_{\ell3}$ decays provide ideal channels for the determination of \Vus.   
The starting point of the analysis is the expression for the
photon-inclusive $K\to\pi\ell\nu$ ($K_{\ell3(\gamma)}$) decay rate:
\begin{align}
\begin{split}
\Gamma_{K_{\ell 3}} = {} &
\frac{G_F^2m_K^5}{192\pi^3}\, C_K^2 S_{\rm EW} 
\left(\Vus f_+^{K^0 \pi^-}(0) \right)^2 I_{K\ell}\\
& \times \left(1 + \dEM{K\ell} + \dSU{K\pi} \right)^2,
\label{eq:Mkl3}
\end{split}
\end{align}
where $G_F$ is the Fermi constant as determined from muon decays, 
$S_{EW}= 1.0232(3)$ \cite{Sirlin:1977sv,Marciano:1993sh} is 
the short-distance electroweak correction, $C_{K}$ is a Clebsch-Gordan 
coefficient ($1$ for $K^0$ and $1/\sqrt{2}$ for $K^\pm$ decays),
$f_+^{K^0 \pi^-} (0)$ is the $K^0 \to \pi^-$ vector form factor at zero 
momentum transfer, and $I_{K\ell}$ is a phase-space integral that is
sensitive to the momentum dependence of the form factors. 
The latter describe the hadronic matrix elements
\begin{eqnarray}
\langle \pi(p_\pi) | \bar{s}\gamma_{\mu}u | K(p_K)\rangle &=& \nonumber \\
(p_\pi+p_K)_\mu f^{K \pi}_+(t) &+& (p_K-p_\pi)_\mu f_-^{K \pi }(t),         
\label{eq:element}
\end{eqnarray}
where $t=(p_K-p_\pi)^2= (p_\ell+p_\nu)^2$. 
The vector form factor $f_+(t)$ represents the P-wave projection 
of the crossed channel matrix element 
$\langle 0 |\bar{s}\gamma^{\mu}u| K \pi \rangle$.
The scalar form factor $f_0(t)$ describes the S-wave projection, 
and in terms of $f_+(t)$ and $f_-(t)$ reads
\begin{equation}
f_0(t)= f_+(t) + \frac{t}{m_K^2-m_\pi^2} f_-(t).
\label{eq:f0def}
\end{equation}
By construction, $f_0(0)=\fp$. 
Since \fp\ is not directly measurable, 
it is convenient to factor out $f_+^{K^0 \pi^-}(0)$ in \Eq{eq:Mkl3} 
and then normalize the form factors for all 
channels to $f_+^{K^0 \pi^-}(0)$, denoted simply as \fp\ in the following. 
The normalized form factors are then defined as
\begin{equation}
\bar f_+(t) = \frac{f_+(t)}{\fp},
\ \bar f_0(t) = \frac{f_0(t)}{\fp},
\ \bar f_+(0) = \bar f_{0} (0) = 1.
\label{eq:Normff}
\end{equation}
Finally, \dEM{K\ell} represents the channel-dependent 
long-distance EM corrections (\Sec{sec:KlSM}) and 
\dSU{K\pi} the correction for isospin breaking
(\Sec{sec:isobreak}).

To extract \Vus\ from $K_{\ell3}$ decays using \Eq{eq:Mkl3}, 
one must measure one or more photon-inclusive $K_{\ell3}$ decay rates,
compute the phase space integrals from form factor measurements, and 
make use of theoretical results for \fp, \dEM{K\ell}, and \dSU{K\pi}.
We discuss the evaluation of these different ingredients in the following.
\end{sloppypar}

\subsubsection{Theoretical determination of \fp}
\label{sec:fp}

The vector form factor at zero momentum transfer \fp\ is the most critical
hadronic quantity required for the determination of \Vus\ from $K_{\ell 3}$ 
decays via \Eq{eq:Mkl3}.
By construction, \fp\ is defined in the absence of electromagnetic 
corrections. More explicitly, \fp\ is defined 
by the $K^0 \to \pi^-$ matrix element of the vector current, 
\Eq{eq:element}, keeping kaon and pion masses at their physical 
values.

In this section, we restrict our discussion to the evaluation of 
\fp\ in the isospin limit\footnote{The choice 
of the $K^0\to\pi^-$ form factor as the common normalization 
is motivated by its smoothness in the $m_u = m_d$ limit 
(see \Sec{sec:isobreak}).} ($m_u = m_d$). 
This hadronic quantity cannot be computed in perturbative QCD, but
is highly constrained by \SU{3} and chiral symmetry. 
In the chiral limit and, more generally, in the \SU{3} limit
($m_u=m_d=m_s$) the conservation of the vector current 
implies $\fp=1$. Expanding around the chiral limit in powers 
of light quark masses one can write
\begin{equation}
\label{eq:chpt4}
\fp = 1 + f_2 + f_4 + \ldots
\end{equation}
where $f_n = \order{m_{u,d,s}^n/(4\pi\Fpi)^n}$, and 
$f_2$ and $f_4$ are the next-to-leading order (NLO) and 
next-to-next-to-leading order (NNLO) corrections in ChPT. The 
Ademollo-Gatto theorem implies that $[\fp - 1]$ is at least of second order 
in the breaking of \SU{3} or in the expansion in powers of $m_s-\hat m$, 
where $\hat m=(m_u+m_d)/2$. This also implies 
that $f_2$ is free of $\order{p^4}$ counterterms in ChPT 
and can be computed with high accuracy: 
$f_2=-0.0227$ \cite{Gasser:1984ux,Leutwyler:1984je}. 

\begin{figure*}[t]
\begin{center}
\vskip 0.2 cm
\includegraphics[width=0.63\textwidth]{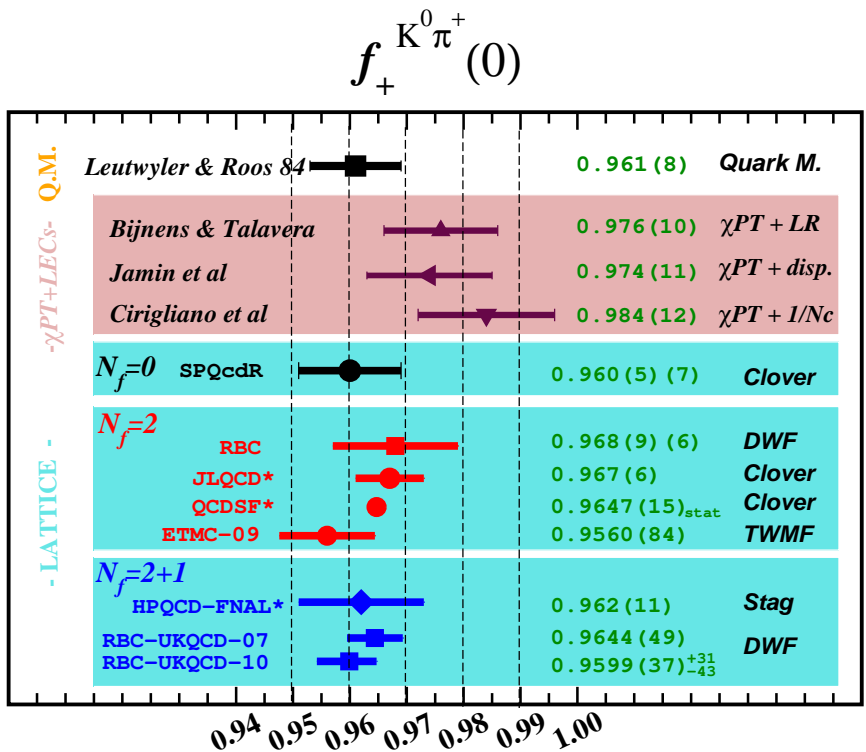}
\vskip 0.5 cm
\caption{\label{fig:f0} Present determinations 
of $\fp \equiv f_+^{K^0\pi^-}(0)$ from analytical or 
semi-analytical approaches \cite{Leutwyler:1984je,Bijnens:2003uy,Jamin:2004re,Cirigliano:2005xn} 
and lattice QCD \cite{Becirevic:2004ya,Tsutsui:2005cj,Dawson:2006qc,Brommel:2007wn,Lubicz:2009ht,Boyle:2007qe,Boyle:2010bh}.
Unpublished results are indicated by asterisks.}
\end{center}
\end{figure*}

\begin{sloppypar}
Difficulties arise with the calculation of the quantity $\Delta f$ 
\begin{equation}
\Delta f \equiv f_4 + f_6 + \ldots = \fp - (1 + f_2),
\end{equation}
which depends on the LECs of ChPT. 
The original quark-model estimate of Leutwyler and Roos \cite{Leutwyler:1984je} 
gives $\Delta f = -0.016(8)$ and $\fp = 0.961 (8)$. 
More recently, analytical calculations have been performed to evaluate the 
NNLO term $f_4$, writing it as  
\begin{equation}
f_4 = \Delta(\mu) + f_4\vert^{\rm loc}(\mu), 
\label{eq:f4ch}
\end{equation}
where $\Delta(\mu)$ is the loop contribution, with $\mu$ the 
renormalization scale, computed in \Ref{Bijnens:2003uy}, 
and $f_4\vert^{\rm loc}(\mu)$ is the $\order{p^6}$ local contribution 
involving $\mathcal{O}(p^4)$ and unknown $\order{p^6}$ LECs. 
To estimate the latter term, various approaches have been used, including
a quark model \cite{Bijnens:2003uy}, dispersion relations \cite{Jamin:2004re}, 
and $1/N_C$ estimates \cite{Cirigliano:2005xn,Kastner:2008ch}.
These studies obtain the results
$\Delta f = 0.001(10)$, $-0.003(11)$, $0.007(12)$, and $0.009(9)$, 
respectively. $\Delta f$ is found to be compatible with zero from these 
studies.
Relative to the quark-model estimate from \Ref{Leutwyler:1984je},
these new results are obtained using more sophisticated techniques
and feature better control of systematic uncertainties.
The resulting values for \fp\ are summarized in \Fig{fig:f0}.
As can be seen, the uncertainties are no smaller 
that those from the original estimate of \Ref{Leutwyler:1984je},
which illustrates the difficulty of calculating $\Delta f$ to below the 
1\% level using analytical methods only.
\end{sloppypar}

\Figure{fig:f0} also summarizes present results for \fp\ from lattice 
QCD \cite{Becirevic:2004ya,Tsutsui:2005cj,Dawson:2006qc,Brommel:2007wn,Lubicz:2009ht,Boyle:2007qe,Boyle:2010bh}. 
As can be seen in \Fig{fig:f0}, lattice results, while in 
remarkable agreement with the original quark-model estimate of 
\Ref{Leutwyler:1984je}, give somewhat smaller results for \fp\
than do recent analytical calculations.

On the lattice, \fp\ is determined using three-point functions at
non-vanishing momenta, whereas \FKpi\ is determined from two-point
functions at rest. Because of this additional difficulty, current 
lattice calculations of \fp\ are less mature than those of \FKpi.
In particular, most results shown in \Fig{fig:f0} were obtained 
with only one lattice spacing and with heavy pion masses. 
Additionally, only one calculation of \fp\ exists with \NF{2+1}: that from
RBC/UKQCD \cite{Boyle:2007qe,Boyle:2010bh}. 
While BMW, MILC, and HPQCD currently have interesting results for \FKpi,
these groups do not yet have results for \fp.

\begin{sloppypar}
Nevertheless, the special chiral properties of \fp\ make
it possible to obtain lattice estimates with relative uncertainties 
comparable to those for \FKpi.
Among the lattice results for \fp\ in \Fig{fig:f0}, 
particularly noteworthy are the studies from 
RBC/UKQCD~'07 \cite{Boyle:2007qe}, RBC/UKQCD~'10 \cite{Boyle:2010bh}, and
ETMC \cite{Lubicz:2009ht}. The corresponding values of \fp\ are
respectively 
\begin{subequations}
\label{eq:f0}
\begin{align} 
\fp &= 0.9644(33)\stat(34)_{\rm syst\mhyp extr}(14)_{\rm syst\mhyp lat}, \\
\fp &= 0.9599(33)\stat(^{+31}_{-43})_{\rm syst\mhyp extr}(14)_{\rm syst\mhyp lat}, \\
\fp &= 0.9560(57)\stat(62)\syst.
\end{align}
\end{subequations}
(The two contributions to the systematic errors
on the RBC/UKQCD results are from extrapolation uncertainties 
and lattice effects, respectively.)
In \Ref{Boyle:2007qe} and in their update \cite{Boyle:2010bh},
RBC/UKQCD make use of a simulation with \NF{2+1}, but a rather coarse 
lattice spacing ($a=0.11$~fm) and a lightest pion mass of 
$m_\pi\approx 330$~MeV. 
Even though the use of smaller lattice spacings would be advisable, the 
corresponding error on the \SU{3} breaking of \fp\ seems to be 
under control to within the stated systematic uncertainty (see, e.g.,
discussion in \Ref{Lubicz:2010nx}).
A critical issue is the chiral extrapolation
from their points with $m_\pi\gtrsim 330$ MeV.
In the new study from RBC/UKQCD \cite{Boyle:2010bh}, there is an attempt 
to quantify the systematic error from the extrapolation, but a
better understanding of the NNLO terms in the chiral
expansion,\footnote{\Ref{Bernard:2009ds} describes exploratory NNLO fits 
using RBC/UKQCD results for $f_0(t)$.}
as well as additional \NF{2+1} simulation points at smaller pion
masses, are needed to fully address this delicate point.
\end{sloppypar}

The recent ETMC study \cite{Lubicz:2009ht} makes use of very light pions,
as well as different lattice spacings and volumes.
Both \SU{2} and \SU{3} chiral fits are investigated and give compatible 
results, which is in contrast to the findings of RBC/UKQCD, in which 
the \SU{2} chiral extrapolation for \fp\ has few points with 
$m_\pi \leq 400$~MeV and looks problematic \cite{Boyle:2010bh}. 
In summary, the ETMC result features a more thorough exploration of systematics
than the RBC/UKQCD result. However, the ETMC simulation still has \NF{2},
and the final uncertainty on \fp\ is inflated to take into account unknown 
systematics related to the quenching of the strange quark.
At present, many other groups are occupied with finalizing their studies of
\FKpi. As attention returns to \fp, further progress on understanding these
systematics should be possible.

For the numerical analysis in Section~\ref{sec:res} we use 
as our reference number 
\begin{equation}
\fp = 0.959(5),
\label{eq:fpQCD}
\end{equation}
which is our symmetrization of the recent RBC/UKQCD result \cite{Boyle:2010bh}.
However, we note that this value is fairly representative of the results 
and spread of values in \Eqs{eq:f0}.
This number is also basically consistent with the average of the 
\NF{2} ETMC and \NF{2+1} RBC/UKQCD results as quoted at the most recent
Lattice conference $\fp=0.962(5)$ \cite{Lubicz:2010nx}.

\subsubsection{Electromagnetic effects in $K_{\ell3}$ decays}
\label{sec:KlSM}

\begin{sloppypar}
The results of the most recent calculation \cite{Cirigliano:2008wn} of the
long-distance electromagnetic corrections to the fully-inclusive
$K_{\ell3(\gamma)}$ decay rates for each of the four modes (the \dEM{K\ell}
in \Eq{eq:Mkl3}) are listed in \Tab{tab:Kl3radcorr}.
These values were obtained to leading nontrivial order in  
chiral effective theory, working with a fully-inclusive prescription of 
real photon emission. For the low-energy electromagnetic couplings 
appearing in the structure-dependent contributions, the recent determinations 
of \Refs{Ananthanarayan:2004qk} and~\citen{DescotesGenon:2005pw} were used.
\begin{table}[t]
\centering
\begin{tabular}{cc}
\hline\hline
Mode & $\dEM{K\ell}$ (\%) \\
\hline 
$K^0_{e3}$      &  0.495 $\pm$ 0.110 \\
$K^\pm_{e3}$    &  0.050 $\pm$ 0.125 \\
$K^0_{\mu3}$    &  0.700 $\pm$ 0.110 \\
$K^\pm_{\mu3}$  &  0.008 $\pm$ 0.125 \\
\hline\hline
\end{tabular}
\caption{Electromagnetic corrections to the fully-inclusive 
$K_{\ell 3(\gamma)}$ rate \cite{Cirigliano:2008wn}.} 
\label{tab:Kl3radcorr}
\end{table}
The errors in \Tab{tab:Kl3radcorr} are estimates of higher-order
contributions that are only partially known. The associated correlation
matrix was found to be \cite{Cirigliano:2008wn}
\begin{equation}
\left(
\begin{array}{rrrr}
+1.000 & +0.081 & +0.685 & -0.147 \\
       & +1.000 & -0.147 & +0.764 \\
       &        & +1.000 & +0.081 \\
       &        &        & +1.000 
\end{array} 
\right).
\end{equation}
It is also useful to list the uncertainties on the linear combinations 
of \dEM{K\ell} that are relevant for lepton-universality and 
strong isospin-breaking tests (as in \Secs{sec:ib} and~\ref{sec:LU}):
\begin{subequations}
\begin{align}
\dEM{K^0 e} - \dEM{K^0 \mu}     &= (-0.205 \pm 0.085)\%, \\
\dEM{K^\pm e} - \dEM{K^\pm\mu}  &= (0.042 \pm 0.085)\%, \\
\dEM{K^\pm e} - \dEM{K^0 e}     &= (-0.445 \pm 0.160)\%, \\
\dEM{K^\pm \mu} - \dEM{K^0 \mu} &= (-0.692 \pm 0.160)\%.
\end{align}
\end{subequations}
The corresponding electromagnetic corrections to the Dalitz plot densities 
can be found in \Ref{Cirigliano:2008wn}. 
It is important to note that the 
corrections to the Dalitz distributions can be locally large (up to 
$ \sim 10 \%$), with considerable cancellations in the integrated 
electromagnetic corrections. Hence a proper implementation of 
the electromagnetic corrections in the analysis of 
the experimental data is essential, in particular for a reliable  
extraction of the form factor parameters. 
\end{sloppypar}

\subsubsection{Isospin-breaking corrections in $K_{\ell3}$ decays}
\label{sec:isobreak}

\begin{sloppypar}
In \Eq{eq:Mkl3}, the same quantity $\fp \equiv f_+^{K^0 \pi^-}(0)$ 
(form factor at zero momentum transfer) is factored out for all
decay channels. The isospin-breaking corrections are then 
included via the term containing \dSU{K\pi}, where
\begin{equation}
\dSU{K^0\pi^-} = 0, \quad
\dSU{K^+\pi^0} = 
\frac{f_+^{K^+\pi^0}(0)}{f_+^{K^0\pi^-}(0)} - 1.
\end{equation}
This term can be related to the $\pi^0$-$\eta$ 
mixing \cite{Cirigliano:2001mk,Kastner:2008ch}
At leading order (\order{p^2}) \cite{Gasser:1984ux}
\begin{equation}
\dSU{K^+\pi^0} = \frac{3}{4} \frac{1}{R},\ \mathrm{with}\ R =  
\frac{m_s - \hat{m}}{m_d - m_u}, 
\label{eq:R} 
\end{equation} 
while at NLO in the chiral expansion (\order{p^4}) \cite{Cirigliano:2007zz}
\begin{equation}
\dSU{K^+\pi^0} = \frac{3}{4} 
\frac{1}{R} \left(1 + \chi_{p^4} 
+ \Delta_M + \order{m_q^2}\right),
\label{eq:Rextrac} 
\end{equation}
where $\chi_{p^4} \approx 0.219$ is an \order{p^4} 
correction calculable in ChPT \cite{Gasser:1984ux}.
$\Delta_M$ is a correction (starting at \order{m_q})
to the ratio $m_K^2/m_\pi^2$: 
\begin{equation}
\frac{m_K^2}{m_\pi^2} = \frac{1}{2} \left(1 + \frac{m_s}{\hat{m}}\right) 
\left(1 + \Delta_M\right) = \frac{Q^2}{R} \left(1 + \Delta_M\right),
\label{eq:dm}
\end{equation}
where $Q^2 = (m_s^2 - \hat{m}^2)/(m_d^2 - m_u^2)$.
Using \Eq{eq:dm}, \Eq{eq:Rextrac} can be rewritten 
\begin{equation}
\dSU{K^+\pi^0}= \frac{3}{4}\frac{1}{Q^2}
\left[\frac{m_K^2}{m_{\pi}^2} + \frac{\chi_{p^4}}{2} 
\left(1 + \frac{m_s}{\hat{m}}\right)\right],
\label{eq:Rextrac2} 
\end{equation}
which shows how \dSU{K^+\pi^0} is essentially determined by the 
double ratio $Q^2$ (the dependence on $m_s/\hat{m}$ is suppressed 
by the smallness of $\chi_{p^4}$). 
One can extract $Q^2$ from the analysis of the decay 
$\eta \to 3\pi$ or from the kaon mass splitting. A recent analysis 
using the latter method gives \cite{Kastner:2008ch}
\begin{equation}
Q = 20.7 \pm 1.2,  
\end{equation}
and thus (using  $m_s/\hat{m} = 24.7 \pm 1.1$ and including \order{e^2p^2, p^2}
corrections \cite{Cirigliano:2001mk} to \Eq{eq:Rextrac2})
\begin{equation}
\dSU{K^+ \pi^0} = 0.029 \pm 0.004.
\label{eq:isobrk}
\end{equation}
This is based on an evaluation of the low-energy electromagnetic couplings 
\cite{Ananthanarayan:2004qk} leading to a large deviation of 
Dashen's limit \cite{Dashen:1969eg}.  
Note that previous analyses of $\eta \to 3\pi$ decays \cite{Kambor:1995yc} 
give higher results for $Q$, and hence central values for \dSU{K^+ \pi^0} 
below the lower edge of the range of values quoted in \Eq{eq:isobrk}. 
New analyses of this decay based on recent data \cite{Ambrosino:2008ht} 
are in progress \cite{Colangelo:2009db,Kampf} and should shed light 
on this issue. 
\end{sloppypar}

As a final note, the precision reached in the measurement of the
$K_{\ell3}$ decay rates and in the determination of the corrections \dEM{K\ell} 
allow \dSU{K^+\pi^0} to be determined directly 
from data, as discussed in \Sec{sec:ib}.
By means of \Eqs{eq:Rextrac} and~\andEq{eq:Rextrac2}, the empirical
determination of \dSU{K^+\pi^0} can then be used to derive interesting
constraints on the quark mass ratios.

\subsubsection{Parameterization of the form factors}
\label{sec:FFpar}

The last ingredient for the determination of \Vus\ from \Eq{eq:Mkl3} is 
the calculation of the phase space integrals, $I_{K\ell}$
\begin{align}
\begin{split}
I_{K\ell} = \int_{m^2_\ell}^{t_{\rm max}}\!\! & dt\,\frac{1}{m_K^8}\,\lambda^{3/2} 
\left(1+\frac{m^2_\ell}{2t}\right) \left(1-\frac{m^2_\ell}{2t}\right)^2 \\
& \times \left(\bar f^2_+(t) + 
\frac{3m^2_\ell \Delta_{K\pi}^2}{(2t + m^2_\ell)\lambda}\,\bar f_0^2(t)\right), 
\label{eq:IK}
\end{split}
\end{align}
with $\Delta_{K\pi}= m_K^2 - m_\pi^2$, 
$\lambda=[t-(m_K + m_\pi)^2][t-(m_K - m_\pi)^2]$,
and $t_{\rm max} = (m_K - m_\pi)^2$.
In order to calculate the integrals, knowledge is required of the normalized 
vector and scalar form factors defined in \Eq{eq:Normff}.
The form factors can be determined from fits to the measured 
distributions of the $K_{\ell 3}$ decays in $t$ or some equivalent variable
using a given parameterization for the form factors. 

Among the different parameterizations proposed in the literature, 
one can distinguish two classes \cite{Abouzaid:2009ry}. 
Parameterizations based on a systematic mathematical expansion are to date 
the most widely used. 
In this class (Class II by the nomenclature of \Ref{Abouzaid:2009ry}),
one finds the Taylor expansion
\begin{align}
\begin{split}
\bar f_{+,0}^{\rm Taylor}(t) = 1 & + 
\lambda'_{+,0} \frac{t}{m_{\pi^\pm}^2} + 
\frac{1}{2}\lambda''_{+,0} \left(\frac{t}{m_{\pi^\pm}^2}\right)^2 \\
& + \frac{1}{6}\lambda'''_{+,0} \left(\frac{t}{m_{\pi^\pm}^2}\right)^3 + 
\cdots,
\label{eq:Taylor}
\end{split}
\end{align}
where $\lambda'_{+,0}$ and $\lambda''_{+,0}$ are the slope and 
the curvature of the form factors, respectively.
Another Class-II parameterization is the so-called $z$-parameterization 
of \Ref{Hill:2006bq}.

In Class-II parameterizations, the parameters describing the higher order 
terms of the form factor expansion are free to be determined from data.
In practice, this additional freedom greatly complicates the use of such
parameterizations. As noted in \Ref{Franzini:2007zz}, if a quadratic 
parameterization is used for both the vector and scalar terms, fits to
experimental data will provide no sensitivity to $\lambda''_0$ because of 
the strong parameter correlations, especially between $\lambda_0'$ and
$\lambda_0''$. For this reason, existing power-series fits use a 
parameterization in $\lambda_+'$, $\lambda''_+ $, and $\lambda_0$ 
(see \Eq{eq:Taylor}). 
It has been shown in \Ref{Bernard:2006gy} that in order to describe 
the form factor shapes accurately in the physical region, 
one has to go at least up to the second order in the Taylor expansion.
This is quantified in \Ref{Franzini:2007zz}: if the same $K_{\mu3}$ spectrum is 
fitted using both the linear ($\lambda_0$) and quadratic ($\lambda_0'$, 
$\lambda_0''$) parameterizations, one typically finds $\lambda_0 \approx
\lambda_0' + 3\lambda_0''$. Ignoring the quadratic term increases the phase 
space integral by about 0.15\%.
In addition, as discussed below and in \Sec{sec:CTtest}, for tests of
low-energy dynamics involving the Callan-Treiman theorem, $\bar f_0(t)$
must be extrapolated to $t=\Delta_{K\pi} \equiv m_K^2-m_\pi^2$, which is
well above the endpoint of the physical region 
in $t$ for $K_{\mu3}$ decays. A parameterization
that accounts for even higher-order terms is therefore desirable.

The parameterizations belonging to Class I circumvent these problems 
by incorporating additional physical constraints to reduce the number 
of independent parameters. A typical example is the pole parameterization 
\begin{equation}
\bar f_{+,0}^{\rm pole}(t) = \frac{M_{V,S}^2}{M_{V,S}^2-t},
\label{eq:pole}
\end{equation}
where the dominance of a single resonance is assumed, and the corresponding 
pole mass $M_{V,S}$ is the only free parameter. 
While for the vector form factor, a pole parameterization with the dominance of 
the $K^*(892)$ ($M_V \sim 892$ MeV) is in good agreement with the data, 
for the scalar form factor, there is no such obvious dominance. 

The most interesting and well-motivated parameterizations
of Class I are those based on dispersion relations. 
These are based on the observation that the vector and scalar form 
factors are analytic functions in the complex $t$-plane, 
except for a cut along the positive real axis for 
$t \geq t_{\rm lim} \equiv (m_K+m_\pi)^2$, where they develop discontinuities. 
One can therefore write
\begin{equation}
\label{dis}
\bar f_{+,0}(t) = \frac{1}{\pi} \int_{t_{\rm lim}}^\infty\!\! ds'\,
\frac{\Im\,\bar f_{+,0}(s')}{(s'-t-i\epsilon)} + {\rm subtractions},
\end{equation}
where the imaginary part, $\Im\,\bar f_{+,0}(s')$,
can be determined from data on $K\pi$ scattering, 
and the ultraviolet component of the integral is 
absorbed into the (polynomial) subtraction terms.
In the vector case, 
the dispersive parameterization turns out to be numerically very 
similar to the pole parameterization 
due to dominant contribution to $\Im\,\bar f_{+}(s')$ from the $K^*(892)$.
On the other hand, the dispersive parameterization 
is particularly useful in the scalar case, 
where there is no dominant one-particle intermediate state. 

In addition to the analyticity constraints, the scalar form factor must 
satisfy an additional theoretical constraint dictated by chiral symmetry.
The Callan-Treiman (CT) theorem \cite{Callan:1966hu}
implies that the scalar form factor at $t = \Delta_{K \pi} \equiv m_K^2-m_\pi^2$
is determined in terms of \FKpi\ and \fp\ up to \order{m_{u,d}}
corrections:
\begin{equation}
C \equiv \bar f_0(\Delta_{K\pi})=\frac{\FK}{\Fpi}\frac{1}{\fp}+ \Delta_{CT}.
\label{eq:CTrel}
\end{equation}
The quantity $\Delta_{CT} = \order{m_{u,d}/4\pi\Fpi}$ can be evaluated 
in ChPT. At NLO in the isospin limit \cite{Gasser:1984ux}, 
\begin{equation}
\label{eq:DeltaCT}
\Delta_{CT} = \SN{(-3.5\pm 8)}{-3},
\end{equation}
where the error is a conservative estimate of the higher-order
corrections \cite{Leutwyler}. 
Results consistent with \Eq{eq:DeltaCT} from NNLO estimates beyond 
the isospin limit have been presented in \Ref{Kastner:2008ch,Bijnens:2007xa}.
As discussed in \Sec{sec:CTtest}, \Eq{eq:CTrel} provides
a useful test of the consistency of 
the lattice results for \FKpi\ and \fp\
with experimental data on the scalar form factors.

\subsubsection{Dispersive parameterization for the form factors}
\label{sec:disp}

Motivated by the existence of the CT theorem, a particularly appealing 
dispersive parameterization for the scalar form factor 
has been proposed \cite{Bernard:2006gy}. 
Two subtractions are performed, one at $t=0$, where by definition 
$\bar f_{0}(0)=1$, and the other at the CT point, $t = \Delta_{K\pi}$. 
Assuming that the scalar form factor has no zeroes, this leads to 
\begin{equation}
\bar f_0^{\rm disp}(t)=\exp\left[\frac{t}{\Delta_{K\pi}}(\lnC- G(t))\right], 
\label{eq:Dispf}
\end{equation}
with
\begin{align}
\begin{split}
G(t) = & \frac{\Delta_{K\pi}(\Delta_{K\pi}-t)}{\pi} \\ 
& \times \int_{t_{\rm lim}}^{\infty}\!
\frac{ds}{s} \frac{\phi_0(s)}{(s-\Delta_{K\pi})(s-t-i\epsilon)}.
\label{eq:G}
\end{split}
\end{align} 
With this parameterization, the only free parameter 
to be determined from data is $C$. 

\begin{sloppypar}
The phase $\phi_0(s)$ can be identified in the elastic region with the S-wave
$(K\pi)_{I=1/2}$ scattering phase: performing two subtractions
minimizes the contributions from the unknown high-energy phase,
which in \Ref{Bernard:2006gy} is simply and conservatively 
assumed to lie within the interval $[0, 2\pi)$. 
The resulting function $G(t)$ in \Eq{eq:G} does not exceed 20\% of the 
expected value of \lnC, while the
theoretical uncertainties are at most 10\% of the value of
$G(t)$ \cite{Bernard:2006gy}. The expressions for the leading slope
parameters in the Taylor expansion as functions of \lnC\ are
\cite{Bernard:2009zm}
\begin{subequations}
\begin{align}
\lambda'_0 &= \frac{m_\pi^2}{\Delta_{K\pi}}\left[\lnC - G(0)\right], \\
\lambda''_0 &= (\lambda'_0)^2  - 2\frac{m^4_\pi}{\Delta_{K \pi}}G'(0), \\
\lambda'''_0 &= (\lambda'_0)^3 - 6\frac{m^4_\pi}{\Delta_{K \pi}}G'(0)\lambda'_0 
- 3\frac{m_\pi^6}{\Delta_{K \pi}}G''(0),
\end{align}  
\end{subequations}
where
\begin{subequations}
\begin{align}
G(0) &= 0.0398(44), \\
-2\frac{m_\pi^4}{\Delta_{K\pi}}\,G'(0) &= \SN{4.16(56)}{-4}, \\
-3\frac{m_\pi^6}{\Delta_{K\pi}}\,G''(0) &= \SN{2.72(21)}{-5}.
\end{align}
\end{subequations}
\end{sloppypar}

A dispersive representation for the vector form factor can be been built in a 
similar way \cite{Bernard:2009zm}. 
Since there is no equivalent of the CT theorem in this case, 
the two subtractions are both performed at $t=0$. 
The expression analogous to \Eq{eq:Dispf} for the vector form factor is 
\begin{equation}
\bar f_+^{\rm disp}(t) = 
\exp\left[\frac{t}{m_\pi^2}\left(\Lambda_+ + H(t)\right)\right],
\label{eq:Dispfp}
\end{equation}
with
\begin{equation}
H(t)=\frac{m_\pi^2 t}{\pi} \int_{t_{\rm lim}}^{\infty}
\frac{ds}{s^2}
\frac{\phi_+ (s)}
{(s-t-i\epsilon)}. 
\label{eq:H}
\end{equation}
Here the fit parameter is 
$\Lambda_+ \equiv m_\pi^2\,d\!\bar f_+(t)/dt|_{t=0}$ and the phase 
$\phi_+(s)$ is derived from P-wave $(K\pi)_{I=1/2}$ elastic scattering.
As in the case of the scalar form factor, the uncertainty on $H(t)$ has a 
small influence on the determination of $\Lambda_+$. 
The expressions for the leading slopes in the Taylor expansion as functions of 
$\Lambda_+$ are \cite{Bernard:2009zm}.  
\begin{subequations}
\begin{align}
\lambda'_+ &= \Lambda_+, \\
\lambda''_+ &= (\lambda'_+)^2 + 2 m^2_\pi H'(0), \\ 
\lambda'''_+ &= (\lambda'_+)^3 + 6 m^2_\pi H'(0) \lambda'_+ + 3 m^4_\pi H''(0), 
\end{align}
\end{subequations}
where
\begin{subequations}
\begin{align}
2m_\pi^2\,H'(0) &= \SN{5.79(97)}{-4}, \\
3m_\pi^4\,H''(0) &= \SN{2.99(21)}{-5}.
\end{align}
\end{subequations}

\begin{sloppypar}
The principal results presented in the following sections are based on the 
dispersive parameterizations of \Eqs{eq:Dispf} and~\andEq{eq:Dispfp}. 
To evaluate the integrals $I_{K\ell}$ from experimental measurements of the 
parameters \lnC\ and \Lp, we use the polynomial expansion given in
Appendix~\ref{sec:polIk}. 
A detailed discussion of the theoretical uncertainties 
entering into the dispersive parameterization via the functions $G$ and $H$
can be found in \Refs{Bernard:2006gy} and~\citen{Bernard:2009zm}.
\end{sloppypar}

A final remark concerns the isospin-breaking and 
electromagnetic corrections. Throughout this work,   
a universal (isospin-invariant) $t$ dependence is assumed for
the normalized form factors in the absence of electromagnetic effects
(i.e., we neglect strong isospin-breaking effects in the slopes).
Conventionally, the masses appearing in the dispersion integrals 
are chosen to be the charged pion and neutral kaon masses, while 
the correct physical masses are used when evaluating the 
phase space integrals. In principle this is not fully correct.  
For instance, a different correction $\Delta_{CT}$ should be applied 
for neutral and charged decays. However, at present, strong 
isospin-breaking in the slopes can be neglected to well 
within the experimental errors.

\subsection{\mathversion{bold}$K_{\ell3}$ and $K_{\ell2}$ beyond the Standard Model}
\label{sec:bsm}

\subsubsection{Effective Lagrangian for semileptonic decays}

\begin{sloppypar}
The implications of the precision data on $K_{\ell 2}$ and $K_{\ell 3}$ 
decays for SM extensions are most conveniently studied within 
a model-independent effective-theory approach.
Within this framework, the most general set of weak-scale dimension-six 
local operators contributing to the charged-current semileptonic 
transitions is identified in \Ref{Cirigliano:2009wk}, 
under the assumptions that the $\SU{2} \times \U{1}$ symmetry 
is linearly realized, and that in the neutrino sector, only 
left-handed neutrinos appear as weak-scale degrees of freedom.
The resulting low-scale ($\mu \sim \order{1~{\rm GeV}}$)
effective Lagrangian governing the semileptonic transitions
$d^j \to u^i \,  \ell^- \,  \bar{\nu}_\ell$ for a given lepton flavor $\ell$ 
involves five operator structures and reads:
\begin{align}
\begin{split}
{\cal L}_{d^j \to u^i} &= -2\sqrt{2}\,\, G_F^{(0)} V_{ij} \\
        \times \Bigl[ & \left(1 + [v_L]_{ij} \right) \bar \ell_L \gamma_\mu \nu_{\ell L} \, \bar u_L^i \gamma^\mu d_L^j \\ 
                      & + [v_R]_{ij} \ \bar \ell_L \gamma_\mu \nu_{\ell L} \ \bar u_R^i \gamma^\mu d_R^j \\
                      & + [s_L]_{ij} \ \bar \ell_R \nu_{\ell L} \ \bar u_R^i d_L^j + [s_R]_{ij} \ \bar \ell_R \nu_{\ell L} \ \bar u_L^i d_R^j \\
                      & + [t_L]_{ij} \ \bar \ell_R \sigma_{\mu\nu} \nu_{\ell L} \ \bar u_R^i \sigma^{\mu\nu} d_L^j \Bigr] \quad + {\rm h.c.},
\label{eq:leffq}
\end{split}
\end{align}
where $G_F^{(0)}/\sqrt{2} = g^2/(8 m_W^2)$. 
The effective couplings $v_{L,R}$, $s_{L,R}$, and $t_{L}$  
encode information on interactions beyond the SM 
and are of order $v^2/\Lambda^2$, where $v$ is the SM Higgs 
vacuum expectation value and $\Lambda$ is the new physics scale. 
The coupling $v_L$ receives contributions from three gauge-invariant 
weak-scale operators (gauge boson-quark vertex correction, gauge boson-lepton
vertex correction and contact four-fermion) while the other couplings are 
in one-to-one correspondence with gauge-invariant four-fermion operators 
at the weak scale. 
\end{sloppypar}

In general, the effective couplings carry flavor indices  
and the operators considered here contribute to flavor-changing neutral
current (FCNC) processes (this is made explicit by writing 
the operators in $\SU{2}_L$ gauge-invariant form at the weak scale). 
In order to avoid the strong constraints from FCNC, it is convenient to 
classify the operators according to their behavior under the $\U{3}^5$  
flavor symmetry of the SM gauge Lagrangian\footnote{
I.e., the freedom to perform \U{3} transformations in family space
for each of the five fermionic gauge multiplets: $Q_L = (u_L, d_L)$, $u_R$,
$d_R$, $L_L = (\nu_L, e_L)$, $e_R$.} 
and organize the discussion in terms of perturbations around the 
$\U{3}^5$-symmetric limit. 
In practice, we will assume that the underlying TeV scale new physics  
has an approximate $\U{3}^5$ invariance. This can be achieved 
if flavor breaking is suppressed by a mechanism such as 
Minimal Flavor Violation (MFV) \cite{Chivukula:1987py,Hall:1990ac,Buras:2000dm,D'Ambrosio:2002ex,Cirigliano:2005ck} 
or by the hierarchy $\Lambda_{\rm flavor} \gg 1 {\rm TeV}$.

\subsubsection{Phenomenology in the $\U{3}^5$ limit}

We start our discussion by assuming dominance of the $\U{3}^5$-invariant 
operators. These are not constrained by FCNC and can have a relatively low 
effective scale $\Lambda$. Moreover, these operators contribute to a number
of precision electroweak tests. Therefore, with this analysis one can assess
the interplay and relative strength of low-energy charged-current processes
versus other observables (from low-energy to the $Z$ pole). 

\begin{sloppypar}
In the $\U{3}^5$ limit, the phenomenology of charged-current processes is
greatly simplified \cite{Cirigliano:2009wk}:
only the SM operators (proportional to $1 + v_L$) 
survive in the effective Lagrangians both for semileptonic decays 
(\Eq{eq:leffq}) and for muon decay, which process is used to determine 
the Fermi constant.
The effects of new physics can therefore be encoded 
into shifts in the values of the effective Fermi constants for semileptonic 
and muon decay:
\begin{align}
G_F^{\rm SL} &= (G_F)^{(0)} \, \left( 1 +  v_L \right),  \\
G_F^\mu &= (G_F)^{(0)} \, \left(1 + \tilde{v}_L \right),
\end{align}
allowing for different values of $v_L$ in semileptonic and muon decays.
The values of the CKM elements $V_{ij}^{\rm phenom}$ as determined from
semileptonic decays are affected by both these shifts, 
because semileptonic transitions are normalized to the Fermi constant $G_F^\mu$
as determined from muon decay.  
In fact, one has
\begin{equation}
V_{ij}^{\rm phenom} = V_{ij} \, \frac{G_F^{\rm SL}}{G_F^\mu} = 
V_{ij} \, \left( 1 + v_L - \tilde{v}_L \right). 
\label{eq:vphenoFB}
\end{equation}
In the $\U{3}^5$ limit, then, a common shift affects all of the $V_{ij}$ 
(as determined from all channels: vector, axial, etc.), and
the only way to expose contributions from new physics
is to construct universality tests in which the absolute normalization 
of the $V_{ij}$ matters.
For light-quark transitions, this involves checking that the first row 
of the CKM matrix is a vector of unit length. 
Therefore, one is led to define: 
\begin{equation}
\Delta_{\rm CKM} \equiv |V_{ud}^{\rm phenom}|^2 + |V_{us}^{\rm phenom}|^2 
+ |V_{ub}^{\rm phenom}|^2 - 1,
\label{eq:dckm}
\end{equation}
in terms of the $V_{ij}^{\rm phenom}$ determined from semileptonic 
transitions using the standard procedure.
The new-physics contributions to $\Delta_{\rm CKM}$ involve four weak-scale 
gauge-invariant local operators ($\varphi$ denotes the SM Higgs doublet),
\begin{subequations}
\label{eq:operators}
\begin{align}
O_{ll}^{(3)} &= \frac{1}{2} (\overline{L}_L \gamma^\mu \sigma^a L_L) (\overline{L}_L \gamma_\mu \sigma^a L_L), \\
O_{lq}^{(3)} &=  (\overline{L}_L \gamma^\mu \sigma^a L_L) (\overline{Q}_L \gamma_\mu \sigma^a Q_L), \\
O_{\varphi l}^{(3)}&= \! i (\varphi^\dagger D^\mu \sigma^a \varphi)(\overline{L}_L \gamma_\mu \sigma^a L_L)+\!{\rm h.c.},      \\
O_{\varphi q}^{(3)} &= \! i (\varphi^\dagger D^\mu \sigma^a \varphi)(\overline{Q}_L \gamma_\mu \sigma^a Q_L)+\!{\rm h.c.},
\end{align}
\end{subequations}
describing contact four-fermion interactions and gauge boson-fermion 
vertex corrections.
Defining $\hat{\alpha}_i^{(3)} = \eta_i v^2/\Lambda_i^2$ (with $\eta_i = \pm 1$),
one has
\begin{equation}
\Delta_{\rm{CKM}} = 4 \, \left(\hat{\alpha}_{ll}^{(3)} - \hat{\alpha}_{lq}^{(3)} 
- \hat{\alpha}_{\varphi l}^{(3)} + \hat{\alpha}_{\varphi q}^{(3)} \right).
\label{eq:dckmnp}
\end{equation}
In specific SM extensions, the $\hat{\alpha}_i$ are functions of 
the underlying parameters. Therefore, via the above relation one can obtain
the constraints from quark-lepton universality tests on any weakly-coupled SM
extension.
\end{sloppypar}

\begin{sloppypar}
Each of the $\hat{\alpha}$ coefficients also contributes to 
other low- and high-energy precision electroweak observables \cite{Han:2004az}. 
Therefore, we can now address in a model-independent way concrete questions 
such as the following:
\begin{itemize}
\item What is the maximal deviation $|\Delta_{\rm CKM}|$ allowed once
all the precision electroweak constraints have been taken into account?
\item Which observables provide the strongest constraints on the operators
contributing to $\Delta_{\rm CKM}$?
\end{itemize}
In order to quantify the significance of the experimental constraints
on CKM unitarity,
we first calculate the range of $\Delta_{\rm CKM}(\hat{\alpha}_i)$ 
allowed by existing bounds from all precision electroweak 
measurements \cite{Cirigliano:2009wk}. In terms of the best-fit values
and the covariance matrix of the $\hat{\alpha}_i$ \cite{Han:2004az}
obtained from the fit to electroweak precision data, at $90 \%$ C.L. one has
\begin{equation}
\SN{-9.5}{-3} \leq \Delta_{\rm CKM} \leq \SN{0.1}{-3}.
\end{equation}
This result implies that a deviation from CKM unitarity at the level of 
$-1\%$ is not ruled out by precision electroweak tests.  A closer scrutiny 
of the precision data shows that the blame for large deviations of 
$\Delta_{\rm CKM}$ from zero could be attributed almost entirely to 
the operator $O_{lq}^{(3)}$, which is constrained relatively poorly from 
LEP2 hadronic cross section data, 
while the other three operators are severely constrained: 
$O_{ll}^{(3)}$ by the Fermi constant, $O_{\varphi q}^{(3)}$ by hadronic $Z$ 
decays, and $O_{\varphi l}^{(3)}$ by leptonic $Z$ decays.  
\end{sloppypar}

The above discussion implies that even a percent-level test of CKM
unitarity would provide information not available from other 
precision tests at low and high energies.  
Indeed, by \Eq{eq:dckmnp}, a test of CKM unitarity to better 
than one part in $10^3$ 
(e.g., with a $0.5\%$ determination of \Vus\ from kaon decays, 
combined with the $0.02\%$ determination of \Vud\ from nuclear beta decays)
would probe new-physics effective scales $\Lambda$ on the order of 10~TeV. 
As we will show in \Sec{sec:res}, the current level of theoretical and 
experimental precision in the determination of \Vus\ allows this 
prospect to be realized.

\subsubsection{Beyond $\U{3}^5$}
\label{sec:2HDM}

\begin{sloppypar}
Corrections to the $\U{3}^5$ limit can be introduced both within MFV 
and via generic flavor structures. In charged-current processes, 
the main effect of $\U{3}^5$ breaking is to turn on 
the chirality flipping (pseudo)scalar and tensor structures  
in \Eq{eq:leffq}.
In MFV, the coefficients parameterizing deviations from $\U{3}^5$ are 
highly suppressed (the chirality flip is associated with 
insertions of Yukawa matrices).
However, such suppression can be compensated by a corresponding 
$\tan \beta$ enhancement in models with two Higgs doublets of
Type II, such as the Higgs sector of the minimal supersymmetric
extension of the SM (MSSM).\footnote{In such models, 
$\tan \beta = v_2/v_1$ is the ratio of the two Higgs vacuum expectation 
values.}
In this case, the leading non-standard contribution involves 
charged Higgs exchange. To one loop, this generates the coefficient
\begin{equation}
[s_R]_{us} =  -\frac{\tan^2 \beta}{(1 + \epsilon_0 \tan \beta)} 
\frac{m_\ell m_s}{m_{H^+}^2}, 
\end{equation}
where $\epsilon_0$ is a correction factor which is negligible
in the non-supersymmetric case, while it is $\order{1/16\pi^2}$ in the 
MSSM \cite{Isidori:2001fv} (see also \cite{Hou:1992sy,Akeroyd:2003zr}). 
In $K_{\ell2}$ decay, the $H^\pm$-exchange amplitude destructively 
interferes with the SM $W^\pm$ amplitude; for large values of 
$\tan \beta$ ($\sim 50$) and low values of $m_{H^\pm}$ ($\sim$100~GeV), 
the $K_{\ell2}$ rate can be decreased by as much as $5\%$.
\end{sloppypar}

\begin{sloppypar}
A highly sensitive probe of $\U{3}^5$ violating structures
is therefore provided by comparing the value of \Vus\ determined using 
$K_{\mu 2}$ decays, which are helicity suppressed, and $K_{\ell 3}$ decays,
which are helicity allowed.\footnote{In principle, the $H^\pm$ 
exchange amplitude affects $K_{e3}$ and $K_{\mu3}$ decays differently.
For $K_{e3}$, the effect is totally negligible, while for $K_{\mu3}$,
it is substantially smaller than for $K_{\mu2}$, but not totally negligible. 
However, this effect is well below the present theoretical and experimental 
errors.}
In practice,
to minimize the impact of the uncertainties from \FK\ and the  
electromagnetic corrections for $K_{\mu2}$, it is more convenient to 
consider the ratio
\begin{equation}
R_{\mu23} = 
\left(\frac{\FKpi}{\fp}\right)^{\!-1}\!
\left(\left|\frac{V_{us}}{V_{ud}}\right|\frac{\FK}{\Fpi}\!\right)_{\!\mu2}
\frac{\Vud_{0^+\to0^+}}{[\Vusf]_{\ell3}},
\label{eq:Rl23}
\end{equation}
which makes explicit contact with the results of \Secs{sec:vusf} 
and~\ref{sec:fKpVusd}. 
The hadronic uncertainties enter 
through the combination $(\FKpi)/\fp$ and could be reduced if this quantity 
were to be directly computed on the lattice.
Within the SM $R_{\mu23}=R_{\ell23}=1$, while the inclusion of 
Higgs-mediated scalar currents leads to 
\begin{equation}
R_{\mu23} \approx \left|\,1\ -\ \frac{m^2_{K^+}}{m^2_{H^+}}
\frac{\tan^2\beta}{1+\epsilon_0\tan\beta}\,\right|.
\label{eq:2HDM}
\end{equation}
Note that $R_{\mu23}$ also provides interesting constraints on the 
existence of charged right-handed currents that appear at NLO in a 
not-quite-decoupling Higgsless effective theory 
\cite{Bernard:2006gy,Bernard:2007cf}.
\end{sloppypar}

\section{Experimental inputs}
\label{sec:data}

\subsection{Notes on fits to branching-ratio and lifetime measurements}

We perform separate fits to world data on the BRs and lifetimes for the
$K_L$, $K_S$ and $K^\pm$.
The inputs to our fits are the observables actually measured by
each single experiment, such as absolute BRs, ratios of BRs, lifetimes,
or partial widths. For uncorrelated measurements with statistical and
systematic errors quoted separately, we add the errors in quadrature.
In many cases, the results for different quantities measured by the
same experiment have correlated errors. The errors are then described by the
covariance matrix, which must be provided by the experiment. 
The free parameters in our fits are the dominant BRs and the lifetime.
In each case, the BRs are constrained to sum to unity. The penalty
method \cite{Antonelli:2008jg} is used to implement this constraint.
Once a first fit has been performed, scale factors are calculated and 
used as described in the general introduction to the 
Particle Data Group (PDG) compilation \cite{Amsler:2008zzb}.  
The present versions of our fits make use of only published measurements.
Moreover, for a measurement to be included in one of our fits, we require
the following information to be available:
\begin{itemize}
\item for BR measurements, an explicit discussion of a satisfactory 
treatment of radiative corrections, especially when they are of comparable
size to the experimental uncertainty;
\item some details about the estimation of the systematic uncertainties; 
\item for correlated measurements from the same experiment, the complete
covariance matrix.
\end{itemize}

\subsection{\mathversion{bold}Dominant $K_L$ branching ratios and $\tau_{K_L}$}
\label{sec:KL}

Numerous measurements of the principal $K_L$ BRs, or of various ratios
of these BRs, have been published in recent years. 
\begin{table*}
\center
\begin{tabular}{lclcr}
\hline\hline
 Parameter                          & Value         & Source    &  Ref.                     &  \multicolumn{1}{c}{Pull} \\[0.5ex]\hline
 $\tau_{K_L}$                        & 50.92(30) ns  &  KLOE      & \cite{Ambrosino:2005vx}  & $-0.8$\\
 $\tau_{K_L}$                        & 51.54(44) ns  &  Vosburgh  & \cite{Vosburgh:1971zk}   & $+0.9$\\ 
 $\BR{K_{e3}}$                       & 0.4049(21)    &  KLOE      & \cite{Ambrosino:2005ec}  & $-1.3$\\
 $\BR{K_{\mu3}}$                      & 0.2726(16)    &  KLOE       & \cite{Ambrosino:2005ec}  & $+0.5$\\
 $\BR{K_{\mu3}}/\BR{K_{e3}}$          & 0.6640(26)    &  KTeV       & \cite{Alexopoulos:2004sx} & $-1.1$\\
 $\BR{3\pi^0}$                      & 0.2018(24)    &  KLOE      & \cite{Ambrosino:2005ec}  & $+2.4$\\ 
 $\BR{3\pi^0}/\BR{K_{e3}}$          & 0.4782(55)    &  KTeV      & \cite{Alexopoulos:2004sx} & $-0.5$\\
 $\BR{\pi^+\pi^-\pi^0}$             & 0.1276(15)    &  KLOE      & \cite{Ambrosino:2005ec}  & $+1.0$\\ 
 $\BR{\pi^+\pi^-\pi^0}/\BR{K_{e3}}$ & 0.3078(18)    &  KTeV        & \cite{Alexopoulos:2004sx} & $-0.8$\\
 $\BR{\pi^+\pi^-}/\BR{K_{e3}}$      & 0.004856(29)  &  KTeV       & \cite{Alexopoulos:2004sx} & $+0.3$\\
 $\BR{\pi^+\pi^-}/\BR{K_{e3}}$      & 0.004826(27)  &  NA48       & \cite{Lai:2006cf}         & $-0.8$\\
 $\BR{\pi^+\pi^-}/\BR{K_{\mu3}}$    & 0.007275(68)  &  KLOE        & \cite{Ambrosino:2006up}   & $-1.5$\\
 $\BR{K_{e3}}/\BR{\mbox{2 tracks}}$ & 0.4978(35)    &  NA48       & \cite{Lai:2004bt}         & $-0.8$\\
 $\BR{\pi^0\pi^0}/\BR{3\pi^0}$      & 0.004446(25)  &  KTeV       & \cite{Alexopoulos:2004sx} & $+0.6$\\
 $\BR{\pi^0\pi^0}/\BR{\pi^+\pi^-} $ & 0.4391(13)    &  PDG ETAFIT  & \cite{Amsler:2008zzb}  & $-0.5$\\
 $\BR{\gamma\gamma}/\BR{3\pi^0}$    & 0.00279(3)    &  KLOE       & \cite{Adinolfi:2003ca}    & $-0.5$\\
 $\BR{\gamma\gamma}/\BR{3\pi^0}$    & 0.00281(2)    &  NA48       & \cite{Lai:2003vc}         & $+0.3$\\
 $\BR{\pi^+\pi^-\gamma}/\BR{\pi^+\pi^-}$           & 0.0208(3)   & KTeV  & \cite{AlaviHarati:2000pt} & $0.0$\\
 $\BR{\pi^+\pi^-\gamma_{\rm DE}}/\BR{\pi^+\pi^-\gamma}$  & 0.689(21)    & KTeV  & \cite{Abouzaid:2006hy}    & $+0.2$\\
 $\BR{\pi^+\pi^-\gamma_{\rm DE}}/\BR{\pi^+\pi^-\gamma}$  & 0.683(11)    & KTeV  & \cite{AlaviHarati:2000pt} & $-0.1$\\
 $\BR{\pi^+\pi^-\gamma_{\rm DE}}/\BR{\pi^+\pi^-\gamma}$  & 0.685(41)    & E731  & \cite{Ramberg:1992yv}     & $0.0$\\
\hline\hline
\end{tabular}
\caption{\label{tab:KLinputs}
Input data used for the fit to $K_L$ BRs and lifetime. For each measurement,
the normalized residual with respect to the results of the fit
is listed in the last column.}
\end{table*}

\begin{sloppypar}
The KTeV experiment has measured five ratios of the partial widths for the 
six main $K_L$ decays from independent samples of $10^5$--$10^6$ events 
collected with a single trigger \cite{Alexopoulos:2004sx}.
The KTeV results for the ratios  
$\BR{K_{\mu3}}/\BR{K_{e3}}$,
$\BR{\pi^+\pi^-\pi^0}/\BR{K_{e3}}$,
$\BR{3\pi^0}/\BR{K_{e3}}$,
$\BR{\pi^+\pi^-}/\BR{K_{e3}}$, and
$\BR{2\pi^0}/\BR{3\pi^0}$, with total uncertainties ranging from 
0.4 to 1.1\%, are listed in \Tab{tab:KLinputs}. 
The six decay modes involved account for more than 99.9\% of the $K_L$ 
width, so KTeV combines the ratios to determine the absolute BRs. 
We instead use 
the five measured ratios in our global fit to $K_L$ BRs and lifetime.
The correlations between the errors are provided by the experiment, 
and are taken into account in our fit.
\end{sloppypar}

NA48 has measured the ratio of the $K_{e3}$ branching ratio to that 
for all $K_L$ decays to final states with two tracks \cite{Lai:2004bt}.
Using a sample of $\SN{8}{7}$ events, they obtain
$\BR{K_{e3}}/\BR{\mbox{2 tracks}} = 0.4978(35)$.

Using a sample of $\SN{13}{6}$ $\phi\to K_L K_S$ decays in which the 
$K_S$ decays to $\pi^+\pi^-$, providing normalization, KLOE has directly 
measured the BRs for the four main $K_L$ decay channels \cite{Ambrosino:2005ec}.
The results depend on the $K_L$ lifetime through the geometrical 
acceptance of the apparatus.
The values listed in \Tab{tab:KLinputs} were obtained using 
$\tau_{K_L}^{(0)} = 51.54$~ns as a reference value, and depend on the 
actual value of the lifetime as 
$d\,{\rm BR}/{\rm BR} = 0.67\,d\tau_{K_L}/\tau_{K_L}$. 
KLOE also reports results obtained using a constraint on the sum of the
four measured BRs to solve for $\tau_{K_L}$, which significantly reduces 
the uncertainties on the BR measurements. In our $K_L$ fit, we use the 
values listed in \Tab{tab:KLinputs} and explicitly include the lifetime
dependence and other experimental correlations; we do not make use of
the value of $\tau_{K_L}$ obtained in \cite{Ambrosino:2005ec}.

KLOE has also measured $\tau_{K_L}$ directly, by fitting the proper decay time
distribution using $K_L\to3\pi^0$ events, for which the reconstruction
efficiency is high and uniform over a fiducial volume of 
$\sim$$0.4\beta\gamma c\tau_{K_L}$.
They obtain $\tau_{K_L}=50.92(30)$~ns \cite{Ambrosino:2005vx}.
We use this value in the fit.

\begin{sloppypar}
There are two recent measurements of 
$\BR{\pi^+\pi^-}/\BR{K_{\ell3}}$,
in addition to the KTeV measurement of 
$\BR{\pi^+\pi^-}/\BR{K_{e3}}$ discussed above.
The KLOE collaboration 
obtains $\BR{\pi^+\pi^-}/\BR{K_{\mu3}} = \SN{7.275(68)}{-3}$
\cite{Ambrosino:2006up},
while NA48 obtains $\BR{\pi^+\pi^-}/\BR{K_{e3}} = \SN{4.826(27)}{-3}$
\cite{Lai:2006cf}. All of these measurements are fully inclusive of inner
bremsstrahlung. The KLOE measurement is fully inclusive of the direct-emission
(DE) component, DE contributes negligibly to the KTeV measurement, and a
residual DE contribution of 0.19\% has been subtracted from the NA48 value
to obtain the number quoted above. 
\end{sloppypar}

We fit the 12 recent measurements listed above, together with nine
additional ratios of the BRs for subdominant decays. The complete 
list of 21 inputs is given in \Tab{tab:KLinputs}. 
As free parameters, our fit has
the seven largest $K_L$ BRs (those to $K_{e3}$, $K_{\mu3}$, $3\pi^0$,
$\pi^+\pi^-\pi^0$, $\pi^+\pi^-$, $\pi^0\pi^0$ and $\gamma\gamma$) and the 
$K_L$ lifetime.
Our definition of $\BR{\pi^+\pi^-}$ is now fully inclusive 
of inner bremsstrahlung (IB), but exclusive of the DE component.
The fit also includes two additional parameters necessary for the 
treatment of the DE component in the radiation-inclusive $\pi^+\pi^-$
decay width: $\BR{\pi^+\pi^-\gamma}$ and 
$\BR{\pi^+\pi^-\gamma_{\rm DE}}$, the branching ratios for decays to 
states with a photon with decay-frame energy $E^*_\gamma > 20$~MeV, 
and with a photon
from DE with $E^*_\gamma > 20$~MeV, respectively. Other parameterizations
are possible, but this one most closely represents the input data set
and conforms to recent PDG usage. With 21 input measurements, 10 free
parameters, and the constraint that the sum of the BRs (except for
$\BR{\pi^+\pi^-\gamma}$, which is entirely included in the sum
of $\BR{\pi^+\pi^-}$ and $\BR{\pi^+\pi^-\gamma_{\rm DE}}$) equal 
unity, we have 12 degrees of freedom. 
The fit results are summarized in \Tab{tab:KLBR}.
The fit gives $\chi^2/{\rm ndf}=19.8/12$ ($P=7.1\%$). 
The normalized residuals with respect to the result of the fit (the pulls) 
for each input are listed in \Tab{tab:KLinputs}.
\begin{table*}
\center
\begin{tabular}{lcc|rrrrrrrrr}
\hline\hline
Parameter & Value & $S$ & \multicolumn{9}{c}{Correlation matrix (\%)}\\
\hline
\BR{K_{e3}} & 0.4056(9) & 1.3                             &$-29$&$-45$&$-30$&$ +6$&$+10$&$-27$&$-27$&$ +8$&$+15$\\
\BR{K_{\mu3}} & 0.2704(10) & 1.5                          &     &$-50$&$  0$&$ -3$&$-10$&$-32$&$-35$&$+13$&$-16$\\
\BR{3\pi^0} & 0.1952(9) & 1.2                             &     &     &$-37$&$ -1$&$ +9$&$+56$&$+63$&$-13$&$+12$\\
\BR{\pi^+\pi^-\pi^0} & 0.1254(6) & 1.3                    &     &     &     &$ -4$&$-16$&$-15$&$-21$&$ -5$&$-20$\\
\BR{\pi^+\pi^-} & \SN{1.967(7)}{-3} & 1.1                 &     &     &     &     &$+14$&$+34$&$ +1$&$ -3$&$+19$\\
\BR{\pi^+\pi^-\gamma}  & \SN{4.15(9)}{-5} & 1.6           &     &     &     &     &     &$+16$&$ +8$&$ -3$&$+74$\\
\BR{\pi^+\pi^-\gamma_{\rm DE}} & \SN{2.84(8)}{-5} & 1.3    &     &     &     &     &     &     &$+35$&$-10$&$+22$\\
\BR{2\pi^0} & \SN{8.65(4)}{-4} & 1.4                      &     &     &     &     &     &     &     &$ -8$&$+10$\\
\BR{\gamma\gamma} & \SN{5.47(4)}{-4} & 1.1                &     &     &     &     &     &     &     &     &$ -4$\\
$\tau_{K_L}$ & 51.16(21)~ns & 1.1                             &     &     &     &     &     &     &     &     &     \\
\hline\hline
\end{tabular}
\caption{Results of fit to $K_L$ BRs and lifetime.}
\label{tab:KLBR}
\end{table*}
\begin{figure*}
\centering
\includegraphics[width=0.8\textwidth]{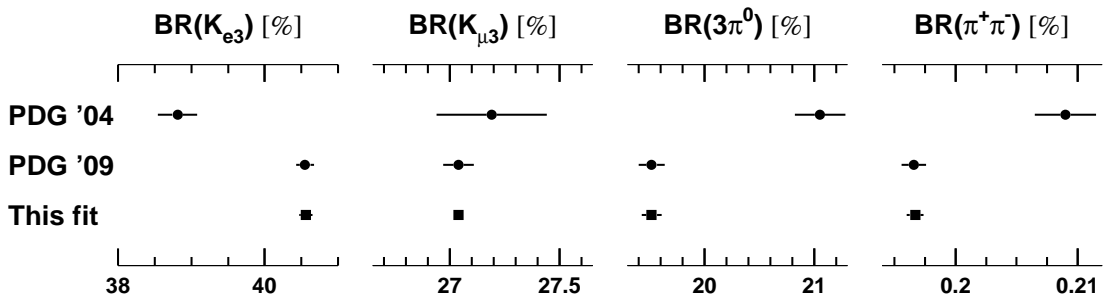}
\caption{Evolution of average values for main $K_L$ BRs.}
\label{fig:klavg}
\end{figure*}
The evolution of the average values of the BRs for $K_{\ell3}$ decays and for
the important normalization channels is shown in \Fig{fig:klavg}.

As seen from \Fig{fig:klavg}, both our fit and the recent PDG fit
\cite{Amsler:2008zzb}
differ substantially in their results from the 2004 PDG fit 
\cite{Eidelman:2004wy}.
This is due to the addition to the world data set after 2004 of most
of the recent data discussed above, and to the subsequent elimination
of numerous old measurements---many with questionable radiative corrections
and/or unreported correlations---used in previous fits.
Between 2004 and today, the world-average BRs for the $K_{e3}$, 
$3\pi^0$, and $\pi^+\pi^-$ decays have shifted by 
$+6\,\sigma$, $-6\,\sigma$, and  $-5\,\sigma$, respectively, 
leading to the following consequences:
\begin{itemize}
\item The world-average value for the ratio $\Gamma(K_{\mu3})/\Gamma(K_{e3})$
      changes from 0.701(8) to 0.6668(28), procuring better agreement 
      with the value expected from lepton universality, as discussed in 
      \Sec{sec:LU}.
\item The world-average value for the amplitude ratio 
      $|\eta_{+-}|$, a parameter of $CP$ violation in the 
      $K_SK_L$ system, 
      changes from \SN{2.286(18)}{-3} to \SN{2.231(10)}{-3},
      a $-2.7\,\sigma$ difference. (For this comparison 
      we are interested mainly in the effect of the new $K_L$ data,
      so we use the results of the fit to $K_S$ rate data in 
      \Tab{tab:KSfit} to evaluate $|\eta_{+-}|$ for both years.) 
\end{itemize}

Figure~\ref{fig:klavg} also shows that differences between the results of
our fit and the most recent PDG fit are minor. These fits differ principally
in the following respects:
\begin{itemize}
\item The PDG fit uses five of the six KTeV values for the dominant $K_L$ BRs,
      while our fit uses the five ratios directly.
\item The PDG fit uses three of the four KLOE values for the dominant $K_L$ BRs
      obtained after imposing the constraint on the sum, as well as the 
      KLOE value of $\tau_{K_L}$ obtained by imposing this constraint. We use
      the original KLOE values with their lifetime dependence and omit the 
      value of $\tau_{K_L}$ obtained from the constraint on the sum.
\item The PDG fit includes \BR{e^+e^-\gamma} as a free parameter, and thus
      makes use of four measurements not used in our fit.
\end{itemize}
Our treatment of the contribution from DE in the $\pi^+\pi^-\gamma$ decay is
the same as that used by the PDG, but we omit the measurement of 
$\BR{\pi^+\pi^-\gamma}/\BR{\pi^+\pi^-}$ from E731 \cite{Ramberg:1992yv},
because while the photon energy cutoff in the numerator is well defined
($E^*_\gamma > 20$~MeV), the requirement used in this measurement for
events in the normalization channel to contain no calorimeter clusters
other than those from the $\pi^+\pi^-$ leads to difficulties of interpretation.

\subsection{\mathversion{bold}Dominant $K_S$ branching ratios and $\tau_{K_S}$}
\label{sec:KS}

\begin{sloppypar}
The KLOE collaboration has measured the ratio 
$\BR{K_S\to\pi e \nu}/\BR{K_S\to\pi^+\pi^-}=\SN{10.19(13)}{-3}$
with 1.3\% precision \cite{Ambrosino:2006si}, making possible an independent
determination of \Vusf\ to better than 0.7\%. In 
\cite{Ambrosino:2006sh}, they combine the above measurement with their
measurement $\BR{K_S\to\pi^+\pi^-}/\BR{K_S\to\pi^0\pi^0} = 2.2459(54)$.
Using the constraint that the $K_S$ BRs must sum to unity and assuming
the universality of lepton couplings, they determine the BRs for the 
$\pi^+\pi^-$, $\pi^0\pi^0$, $K_{e3}$, and $K_{\mu3}$ decays.
\end{sloppypar}

Our fit is an extension of the analysis in \cite{Ambrosino:2006sh}.
We perform a fit to the data on the $K_S$ BRs to $\pi^+\pi^-$, 
$\pi^0\pi^0$, and $K_{e3}$ that uses, in addition to the above 
two measurements:
\begin{itemize}
\item the measurement from NA48, 
$\Gamma(K_S\to\pi e\nu)/\Gamma(K_L\to\pi e\nu)=0.993(34)$ \cite{Batley:2007zzb},
where the denominator is obtained from the results of our $K_L$ fit;
\item the measurements
from NA48, $\tau_{K_S} = 89.598(70)$~ps \cite{Lai:2002kd}, 
and KTeV, $\tau_{K_S} = 89.58(13)$~ps \cite{AlaviHarati:2002ye},
both obtained without the assumption of $CPT$ symmetry; 
\item the result $\BR{K_{\mu3}}/\BR{K_{e3}} = 0.6655(15)$ 
obtained from the assumption of universal lepton couplings, the
values of \Lp\ and \lnC, the parameters of the 
dispersive representation of the vector and scalar form factors,
obtained from our fit to form factor data discussed 
in \Sec{sec:Kl3ff}, and the long-distance 
electromagnetic corrections discussed in~\Sec{sec:Kl3}.
\end{itemize} 
The free parameters are the four BRs listed above plus $\tau_{K_S}$.
With six inputs and one constraint (on the sum of the BRs), the fit
has one degree of freedom and gives $\chi^2/{\rm ndf}=0.015/1$ ($P=90\%$).
The scale factor $S$ is not different from unity for any of the output 
values. The results of the fit are listed in \Tab{tab:KSfit}.
\begin{table}
\center
\begin{tabular}{lc|rrrr}
\hline\hline
Parameter & Value & \multicolumn{4}{c}{Correlation matrix (\%)}\\
\hline
$\BR{\pi^+\pi^-}$ & 0.6920(5)          & $-100$  & $  +4$  & $  +3$  & $  +0$ \\  
$\BR{\pi^0\pi^0}$ & 0.3069(5)          &         & $  -6$  & $  -6$  & $  +0$ \\
$\BR{K_{e3}}$     & \SN{7.05(8)}{-4}   &         &         & $ +98$  & $  +1$ \\
$\BR{K_{\mu3}}$   & \SN{4.69(6)}{-4}   &         &         &         & $  +1$ \\
$\tau_{K_S}$      & 89.59(6) ps        &         &         &         &        \\
\hline\hline 
\end{tabular}
\caption{Results of fit to $K_S$ BRs and lifetime.}
\label{tab:KSfit}
\end{table}

\subsection{\mathversion{bold}Dominant $K^\pm$ branching ratios and $\tau_{K^\pm}$}
\label{sec:Kpm}

\begin{table*}
\center
\begin{tabular}{lclcr}
\hline\hline
  Parameter                                          & Value         & Source   & Ref. & \multicolumn{1}{c}{Pull}  \\[0.5ex]\hline
  $\tau_{K^\pm}$                                     & 12.422(40) ns & Koptev* &\cite{Koptev:1995je} &   $+0.9$ \\
  $\tau_{K^\pm}$                                     & 12.380(16) ns & Ott    &\cite{Ott:1971rs} &       $-0.3$ \\ 
  $\tau_{K^\pm}$                                     & 12.443(38) ns & Fitch & \cite{Fitch:1965zz} &     $+1.5$ \\ 
  $\tau_{K^\pm}$                                     & 12.347(30) ns & KLOE  & \cite{Ambrosino:2007xz} & $-1.2$ \\ 
  $\BR{K_{\mu2}}$                                    & 0.6366(17)   & KLOE & \cite{Ambrosino:2005fw}&    $+1.1$ \\ 
  $\BR{\pi\pi^0}$                                   & 0.2065(9)    & KLOE  &  \cite{Ambrosino:2007xs}&  $+0.5$ \\
  $\BR{\pi\pi^0}/\BR{K_{\mu2}}$                      & 0.3329(48)    & PS183 & \cite{Usher:1992pz} &     $+1.7$ \\ 
  $\BR{\pi\pi^0}/\BR{K_{\mu2}}$                      & 0.3355(57)    & Weissenberg & \cite{Vaisenberg:1976tz} & $+1.9$ \\
  $\BR{\pi\pi^0}/\BR{K_{\mu2}}$                      & 0.3277(65)    & Auerbach & \cite{Auerbach:1974ka} & $+0.5$ \\
  $\BR{K_{e3}}$                                      & 0.04965(53)   & KLOE & \cite{Ambrosino:2007xm} & $-2.1$ \\
  $\BR{K_{e3}}/\BR{\pi\pi^0+K_{\mu3}+\pi\pi^0\pi^0}$  & 0.1962(36)    & BNL-E865 & \cite{Sher:2003fb} &      $-0.3$ \\ 
  $\BR{K_{e3}}/\BR{\pi\pi^0}$                        & 0.2470(10)    & NA48/2 & \cite{Batley:2006cj} &  $+0.6$ \\
  $\BR{K_{\mu3}}$                                    & 0.03233(39)   & KLOE & \cite{Ambrosino:2007xm} & $-3.2$ \\
  $\BR{K_{\mu3}}/\BR{\pi\pi^0}$                      & 0.1637(7)     & NA48/2 & \cite{Batley:2006cj} &  $+1.0$ \\  
  $\BR{K_{\mu3}}/\BR{K_{e3}}$                         & 0.671(11)    & KEK-E246 & \cite{Horie:2001th} & $+0.9$ \\ 
  $\BR{\pi\pi^0\pi^0}$                              & 0.01763(26)   & KLOE & \cite{Aloisio:2003jn} &   $+0.2$ \\ 
  $\BR{\pi\pi^0\pi^0}/\BR{\pi\pi\pi}$               & 0.303(9)      &  Bisi  &\cite{Bisi:1965zz} &     $-0.4$ \\
\hline\hline
\end{tabular}
\caption{
Input data used for the fit to $K^\pm$ BRs and lifetime. For each measurement,
the normalized residual with respect to the results of the fit
is listed in the last column.
The two 1995 values of $\tau_{K^\pm}$ from Koptev et al.\ are averaged with 
$S=1.6$ before being included in the fit as a single value.}
\label{tab:KPMinputs}
\end{table*}
Several recent measurements contribute significant new information on the 
rates for the dominant $K^\pm$ decays.
In addition, we have recently carried out a comprehensive, critical survey of 
the $K^\pm$ data set, which led to the elimination of numerous 
older measurements previously used in the fit. 
The input data used in our current fit to determine the dominant 
$K^\pm$ BRs and lifetime are summarized in \Tab{tab:KPMinputs}. 

\begin{sloppypar}
The 2003 measurement of $\BR{K^+_{e3}}$ by E865 \cite{Sher:2003fb}
was the first of the recent-generation measurements of semileptonic
kaon BRs, and gave a value for \Vus\ consistent with unitarity.
The quantity actually measured was
 $\BR{K^+_{e3}}/(\BR{\pi^+\pi^0}+\BR{K^+_{\mu3}}+\BR{\pi^+\pi^0\pi^0})$,
where one $\pi^0$ in the final state was required to undergo
Dalitz decay. (Throughout the remainder of this section, we use $\pi$
to denote the charged pion when no confusion results.)
\end{sloppypar}

\begin{sloppypar}
In 2007, the NA48/2 collaboration published measurements of the ratios
\BR{K_{e3}}/\BR{\pi\pi^0} and \BR{K_{\mu3}}/\BR{\pi\pi^0}
obtained with simultaneous $K^+$ and $K^-$ beams 
\cite{Batley:2006cj,Batley:2006cj:ERR}.
For each type of $K_{\ell3}$ decay (i.e., to $e$ and $\mu$), about 50k
$K^+$ and 30k $K^-$ decays were collected.
The results for these ratios in \Tab{tab:KPMinputs} are
correlated with $\rho=+0.19$ \cite{NA48private}.
The dominant uncertainties are from sample statistics.
\end{sloppypar}

\begin{sloppypar}
ISTRA+ has also performed a measurement of $\BR{K^-_{e3}}/\BR{\pi\pi^0}$
with 0.6\% precision \cite{Romanovsky:2007qb}. The result, however, has
not been officially published and is therefore not used in our fit.
\end{sloppypar}

KLOE has measured the absolute BRs for the
$K_{e3}$ and $K_{\mu3}$ decays \cite{Ambrosino:2007xm}.
In $\phi\to K^+ K^-$ events, $K^+$ decays into $\mu^+\nu$ or $\pi^+\pi^0$
are used to tag a $K^-$ beam, and vice versa. KLOE performs four
separate measurements for each $K_{\ell3}$ BR, corresponding to the
different combinations of kaon charge and tagging decay.
As in the case of KLOE's absolute BR measurements for the $K_L$, there
is some dependence on the lifetime.
The final values for $\BR{K_{e3}}$ and $\BR{K_{\mu3}}$
in \Tab{tab:KPMinputs} are referred to 
$\tau_{K^\pm} = 12.385$~ns
and depend on the actual value of the lifetime as 
$d\,{\rm BR}/{\rm BR} = -0.45\,d\tau_{K^\pm}/\tau_{K^\pm}$.
The uncertainties are correlated, with $\rho = +0.63$. 

As seen from \Tab{tab:KPMinputs}, KLOE has also measured the absolute BRs 
for the important normalization channels
$K^+\to\pi^+\pi^0$ \cite{Ambrosino:2007xs} 
and $K^+\to\mu^+ \nu$ \cite{Ambrosino:2005fw},
using $K^-\to\mu^- \bar{\nu}$ decays to tag.
Again, our fit takes into account the correlation between these values, as
well as their dependence on the $K^\pm$ lifetime.

One of the primary motives for our critical survey of $K^\pm$ decay 
rate data was the poor consistency of the available lifetime measurements.
The 2007 PDG average value for $\tau_{K^\pm}$, 12.385(25)~ns, is nominally
quite precise. However, the error is scaled by 2.1, and the confidence
level for the average is 0.17\% \cite{Yao:2006px}.

Our survey of the older measurements of $\tau_{K^\pm}$ led to the 
elimination of the 1967 result from Lobkowicz et al.\ 
\cite{Lobkowicz:1969mx}, because the experiment was much more suited for
measuring the difference between $\tau_{K^+}$ and $\tau_{K^-}$ than it 
was for the absolute measurement of either lifetime. 
The stopped-$K^+$ measurement described in the 1995 paper from 
Koptev et al.\ \cite{Koptev:1995je} makes use of the surface-muon technique, 
in which kaons are produced and stopped in the same target. 
The two results obtained using different target materials 
are in marginal agreement, with $S=1.6$. We have been able to identify
neither a reason for the discrepancy, nor a reason for the exclusion of
the measurement, and so we include in our fit the average of the two 
results, with the scale factor applied to the error to reflect their
disagreement.

In 2008, KLOE published a new measurement of $\tau_{K^\pm}$ 
\cite{Ambrosino:2007xz}.
The new KLOE result in \Tab{tab:KPMinputs} is the average of
separate measurements for $K_{\mu2}$-tagged $K^+$ and $K^-$ decays 
using two different techniques for each charge.
In the first technique, the path length of the tagged kaon from 
production to decay was measured using an inclusive sample.
In the second technique, the tagged kaon was required to decay to a 
final state containing a $\pi^0$, and the decay time was measured using 
the photon clusters in the calorimeter.

\begin{table*}
\center
\begin{tabular}{lcc|rrrrrr}
\hline\hline
Parameter & Value & $S$ & \multicolumn{6}{c}{Correlation matrix (\%)} \\
\hline
\BR{K_{\mu2}}      & 63.47(18)\%    & 1.3   & $ -39$  & $ -75$  & $ -33$  & $ -28$  & $ -36$  & $ +12$ \\  
\BR{\pi\pi^0}      & 20.61(8)\%     & 1.1   &         & $ -26$  & $ +61$  & $ +38$  & $ -13$  & $ -11$ \\
\BR{\pi\pi\pi}     &  5.73(16)\%    & 1.2   &         &         & $ -22$  & $ -17$  & $ +36$  & $  -5$ \\
\BR{K_{e3}}        &  5.078(31)\%   & 1.3   &         &         &         & $ +47$  & $ -10$  & $ -13$ \\
\BR{K_{\mu3}}      &  3.359(32)\%   & 1.9   &         &         &         &         & $  -8$  & $  -4$ \\
\BR{\pi\pi^0\pi^0} &  1.757(24)\%   & 1.0   &         &         &         &         &         & $  -1$ \\
$\tau_{K^\pm}$         & 12.384(15)~ns  & 1.2   &         &         &         &         &         &        \\
\hline\hline
\end{tabular}
\caption{\label{tab:KpmBR}
Results of fit to $K^\pm$ BRs and lifetime.}
\end{table*}
\begin{figure*}
\centering
\includegraphics[width=0.7\textwidth]{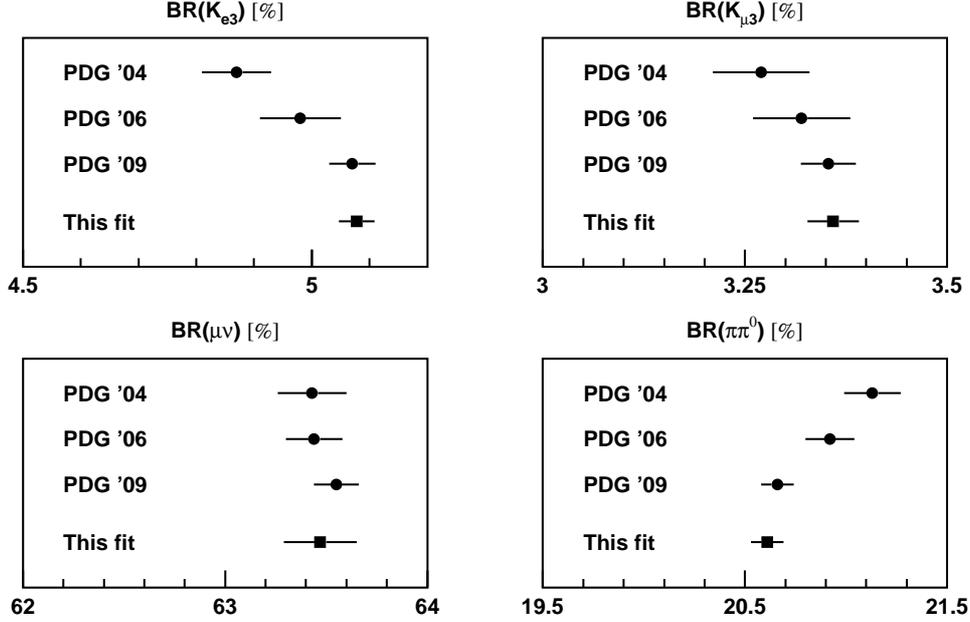}
\caption{Evolution of average values for main $K^\pm$ BRs.}
\label{fig:kpmavg}
\end{figure*}
Our fit for the six largest $K^\pm$ branching ratios and lifetime
uses the 17 measurements in \Tab{tab:KPMinputs}. The fit has seven 
free parameters and one constraint, giving 11 degrees of freedom.
We obtain the results in \Tab{tab:KpmBR}. 
The fit gives $\chi^2/{\rm ndf}=25.8/11$ ($P=0.69\%$). 
The comparatively low $P$-value
reflects some tension between the KLOE and NA48/2 measurements of the 
$K_{\ell3}$ branching ratios.

For comparison, the most recent PDG fit
\cite{Amsler:2008zzb} uses 32 measurements to determine
seven BRs (including that for the $K_{e4}$ decay to $\pi^0\pi^0 e^\pm\nu$)
and the lifetime, and has $\chi^2/{\rm ndf}=52/25$
($P=0.13\%$). The main differences between our fit and the current PDG fit, 
besides the treatment of the $K^\pm$ lifetime data described above, lie 
in the elimination of ten older measurements of various BRs, including the 
five absolute BR measurements of Chiang et al. \cite{Chiang:1972rp}. In that 
study, absolute measurements of the six largest BRs were performed; 
the values were improved by constraining the sum of the six BRs to unity.
Neither the original measurements nor the covariance matrix are reported. 
Moreover, radiative corrections appear not to have been applied when obtaining
the BR values.

Both the significant evolution of the average values of the $K_{\ell3}$
BRs over time and the remaining difference between our results and those
of the PDG are evident in \Fig{fig:kpmavg}.

\subsection{\mathversion{bold}Measurements of $K_{\ell3}$ form factor parameters}

\subsubsection{$K_{e3}$ form factor parameters}

\begin{table}
\center
\begin{tabular}{lccc}
\hline\hline
Experiment & \SN{\lambda_+'}{3} &  \SN{\lambda_+''}{3} & 
$\rho(\lambda_+'$,$\lambda_+'')$ \\
\hline
KLOE & $25.5\pm1.8$ & $1.4\pm0.8$ & $-0.95$ \\
KTeV & $21.67\pm1.99$ & $2.87\pm0.78$ & $-0.97$ \\
NA48 & $28.0\pm2.4$ & $0.4\pm0.9$ & $-0.88$ \\
ISTRA+ & $24.85\pm1.66$ & $1.92\pm0.94$ & $-0.95$ \\
\hline\hline
\end{tabular}
\caption{\label{tab:Ke3ffData}
Quadratic form factor parameters for $K_{e3}$ decays. The values
from NA48 and ISTRA+ have been converted for use with the notation 
of \Eq{eq:Taylor}. The values of $\rho$ from NA48 and ISTRA+ were 
communicated privately \cite{NA48private,ISTRAprivate}.}
\end{table}
\begin{figure}
\centering
\includegraphics[width=0.8\linewidth]{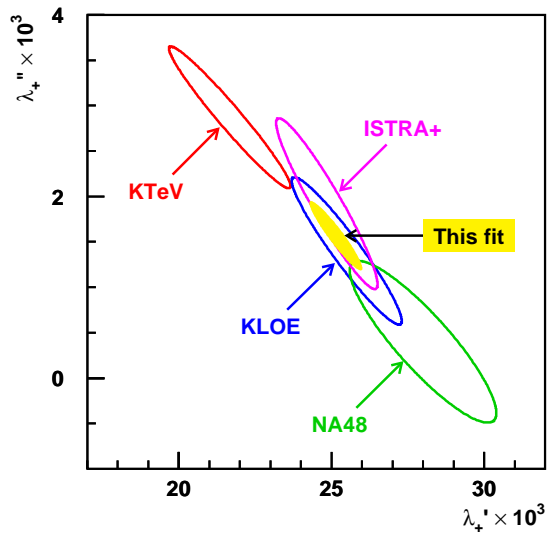}
\caption{Recent measurements of $K_{e3}$ vector form factor parameters. 
The yellow ellipse shows the result of a fit to all data.}
\label{fig:Ke3ff}
\end{figure}
KLOE \cite{Ambrosino:2006gn},
KTeV \cite{Alexopoulos:2004sy},
NA48 \cite{Lai:2004kb}, and
ISTRA+ \cite{Yushchenko:2004zs} have all performed 
recent measurements of the quadratic parameters 
$\lambda_+'$ and $\lambda_+''$
of the vector form factor for $K_{e3}$ decays
(see \Eq{eq:Taylor}). The data are listed in \Tab{tab:Ke3ffData}
and represented graphically in \Fig{fig:Ke3ff}.

\begin{table*}
\center
\begin{tabular}{lcc}
\hline\hline
& $K_L$ and $K^-$ data & $K_L$ data only \\
& 4 measurements & 3 measurements \\
& $\chi^2/{\rm ndf} = 5.3/6$ (51\%) &
  $\chi^2/{\rm ndf} = 4.7/4$ (32\%)\\
\hline
\SN{\lambda_+'}{3}            & $25.1\pm0.9$ & $24.9\pm1.1$ \\
\SN{\lambda_+''}{3}           & $1.6\pm0.4$ & $1.6\pm0.5$ \\
$\rho(\lambda_+',\lambda+'')$ & $-0.94$ & $-0.95$ \\
$I(K^0_{e3})$                  & 0.15463(21) & 0.15454(29) \\
$I(K^\pm_{e3})$                & 0.15900(22) & 0.15890(30) \\
\hline\hline
\end{tabular}
\caption{Average of quadratic fit results for $K_{e3}$ slopes.}
\label{tab:Ke3ff}
\end{table*}
\Tab{tab:Ke3ff} gives the results of a fit to the $K_L$ and $K^-$ data 
in the first column, and to the $K_L$ data only in the second column. 
With correlations taken into account, both fits give good
values of $\chi^2/{\rm ndf}$. The significance of the quadratic
term is $4.2\sigma$ from the fit to all data, and $3.5\sigma$ from
the fit to $K_L$ data only.
Including or excluding the $K^-$ slopes
has little impact on the values of $\lambda_+'$ and $\lambda_+''$;
in particular, the values of the phase space integrals change by just
0.06\%. The errors on the phase space integrals are significantly
smaller when the $K^-$ data are included in the average. 
The results of the fit to all data are plotted as the yellow ellipse in
\Fig{fig:Ke3ff}.
\begin{table*}
\center
\begin{tabular}{lcccc}
\hline\hline
Experiment & $M_V$ (MeV)     & Ref.                      &  \SN{\Lambda_+}{3} & Ref. \\
\hline                     
KLOE       & $870\pm6\pm7$   & \cite{Ambrosino:2006gn}   &  $25.74 \pm 0.53$   & \cite{Ambrosino:2006gn} \\  
KTeV       & $881.03\pm7.11$ & \cite{Alexopoulos:2004sy} &  $25.17 \pm 0.48$   & \cite{Abouzaid:2009ry} \\
NA48       & $859\pm18$      & \cite{Lai:2004kb}         &  $26.4  \pm 1.1$    & \cite{Lai:2004kb} \\        
ISTRA+     & $861.6\pm6.5\pm5.0$ & \cite{ISTRAprivate}       &  $26.20 \pm 0.56$   & \cite{ISTRAprivate} \\      
Average    & $871\pm5$                      &            &  $25.70 \pm 0.42$   &  \\
           & $\chi^2/{\rm ndf} = 3.8/3$     &            & $\chi^2/{\rm ndf} = 2.42/3$ \\
\hline
           & $I(K^0_{e3}) = 0.15480(18)$     &            & $I(K^0_{e3}) = 0.15478(18)$ \\
           & $I(K^\pm_{e3}) = 0.15918(19)$   &            & $I(K^\pm_{e3}) = 0.15924(19)$ \\
\hline\hline
\end{tabular}
\caption{$K_{e3}$ form factor parameters for the pole and dispersive
parameterizations. Values of $\Lambda_+$ from KLOE and NA48 were obtained
from the pole fit results from those experiments.}
\label{tab:Ke3pd}
\end{table*}

\begin{sloppypar}
All four experiments have fitted their $K_{e3}$ data using the
pole parameterization of \Eq{eq:pole}, and 
obtain the values shown in the left part of \Tab{tab:Ke3pd}
for the pole mass $M_V$.
The average value is
$M_V = 871\pm5$~MeV with $\chi^2/{\rm ndf} = 3.8/3$ ($P=28.9\%$).
The individual values are quite compatible with each other, and
their average is reasonably close to the mass of the $K^*(892)$.
The values of the integrals $I(K^0_{e3})$ and $I(K^\pm_{e3})$ as evaluated 
from the pole fit results are just 0.11\% higher than the values obtained
from the quadratic fit results. 
\end{sloppypar}

The dispersive parameterization of the vector form factor of \Eq{eq:H}
is similar to the pole parameterization, but better motivated theoretically.
KTeV and ISTRA+ have used this form to fit their $K_{e3}$ data, while
results for KLOE and NA48 can be obtained from the corresponding values 
of $M_V$ via $\Lambda_+ = (m_{\pi^\pm}/M_V)^2$.
These results are listed in the right part of \Tab{tab:Ke3pd}, 
together with our average.
The uncertainties on the values from the individual experiments do 
not include the contribution arising from the representation
of the form factor phase in the dispersive parameterization
(see discussion in \Ref{Abouzaid:2009ry}).
This contribution is common to all experiments, and is included as an
additional uncertainty of \SN{0.30}{-3} on our average value of $\Lambda_+$,
which is propagated in the evaluation of the 
phase space integrals.

\subsubsection{$K_{\ell3}$ form factor parameters}
\label{sec:Kl3ff}

\begin{table*}
\center
\begin{tabular}{lccccccc}
\hline\hline
Experiment & \SN{\lambda_+'}{3} & \SN{\lambda_+''}{3} & \SN{\lambda_0}{3} & 
  $\rho(\lambda_+',\lambda_+'')$ & $\rho(\lambda_+',\lambda_0)$ & 
  $\rho(\lambda_+'',\lambda_0)$ & Refs. \\
\hline
KTeV &
     $20.64\pm1.75$ & $3.20\pm0.69$ & $13.72\pm1.31$ &
     $-0.97$ & $+0.34$ & $-0.44$ & \cite{Alexopoulos:2004sy} \\
KLOE &
     $25.6\pm1.8$ & $1.5\pm0.8$ & $15.4\pm2.2$ &
     $-0.95$ & $+0.29$ & $-0.38$ & \cite{Ambrosino:2007yz} \\
NA48 & 
     $24.86\pm1.88$ & $1.11\pm0.74$ & $10.25\pm1.05$ & 
     $-0.93$ & $+0.38$ & $-0.51$ & \cite{Lai:2004kb,Lai:2007dx} \\
ISTRA+ &
     $24.80\pm1.56$ & $1.94\pm0.88$ & $16.71\pm1.20$ & 
     $-0.94$ & $+0.34$ & $-0.44$ & \cite{Yushchenko:2004zs,Yushchenko:2003xz,ISTRAprivate} \\ 
\hline
Our avg &
     $25.02\pm1.12$ & $1.57\pm0.48$ & $13.34\pm1.41$ & 
     $-0.950$ & $+0.376$ & $-0.573$ & \\
Our avg, no NA48 $K_{\mu3}$ &
     $25.04\pm0.82$ & $1.57\pm0.36$ & $15.90\pm0.79$ & 
     $-0.942$ & $+0.234$ & $-0.349$ & \\
\hline\hline
\end{tabular}
\caption{
Quadratic-linear form factor parameters for $K_{\ell3}$ decays
from KLOE, KTeV, NA48, and ISTRA+.
For each experiment, the results for $K_{e3}$ and $K_{\mu3}$ decays are
averaged.
The input $K_{e3}$ values from NA48 and all values from ISTRA+ have 
been converted for use with the notation of \Eq{eq:Taylor}; correlation 
coefficients for these measurements were communicated privately.
Our averages of the results from all four experiments, with and 
without the NA48 $K_{\mu3}$ data included, are also listed.}
\label{tab:Kl3ffdata}
\end{table*}
\begin{figure*}
\center
\includegraphics[width=0.8\textwidth]{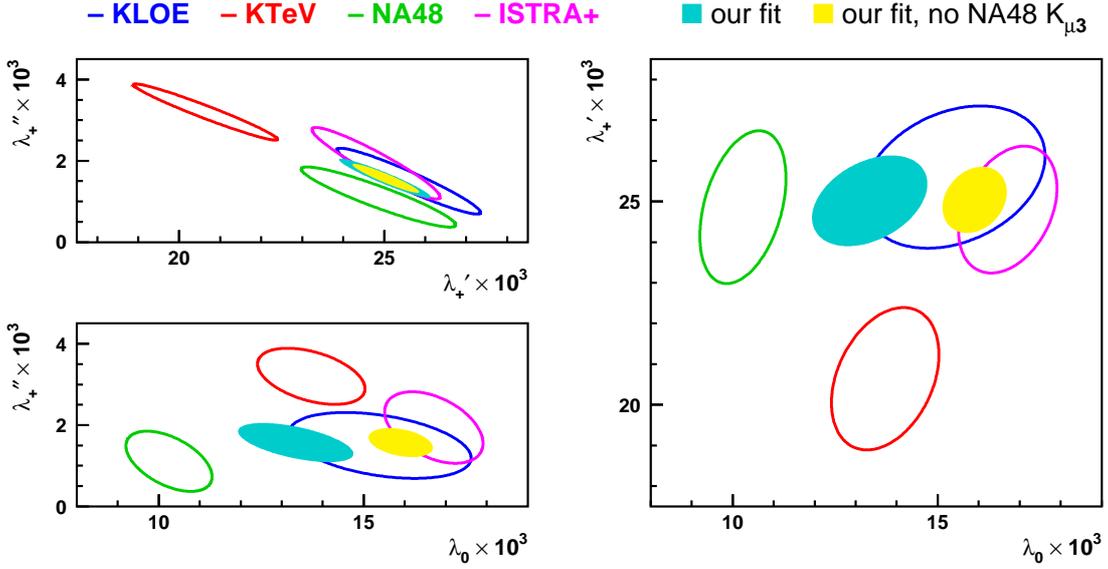}
\caption{$1\sigma$ confidence contours for measurements of 
$\lambda_+'$, $\lambda_+''$, and $\lambda_0$ from KLOE, KTeV, NA48, and ISTRA+.
For each experiment, the results for $K_{e3}$ and $K_{\mu3}$ decays are
averaged.
Our averages of the results from all four experiments, with and 
without the NA48 $K_{\mu3}$ data included, are also shown.}
\label{fig:Kl3ff}
\end{figure*}
KLOE, KTeV, NA48, and ISTRA+ have all performed measurements of the
$K_{\mu3}$ form factor parameters using the quadratic and linear
parameterizations for the vector and scalar form factor parameters, 
respectively. The values of $\lambda_+'$ and $\lambda_+''$ from $K_{e3}$ 
decays may be averaged with the values of $\lambda_+'$, $\lambda_+''$, 
and $\lambda_0$ from $K_{\mu3}$ decays, with parameter correlations taken
into account. KTeV and KLOE provide $K_{e3}$-$K_{\mu3}$ averages for the
slope parameters; we calculate the averages for NA48 and ISTRA+. 
The $K_{e3}$-$K_{\mu3}$ average values for $\lambda_+'$, $\lambda_+''$, 
and $\lambda_0$ from each experiment are listed in \Tab{tab:Kl3ffdata}; the
$1\sigma$ confidence contours for each pair of parameters are plotted 
in \Fig{fig:Kl3ff}. The only significant difference between the results 
presented in \Tab{tab:Kl3ffdata} and \Fig{fig:Kl3ff} and the corresponding
presentation in \Ref{Antonelli:2008jg} is that the final ISTRA+ systematics 
are now correctly included. As has been previously noted 
\cite{Antonelli:2008jg,Antonelli:2009ws}, the NA48 result is incompatible
with all of the other measurements. Our average of the results from all 
four experiments is plotted in \Fig{fig:Kl3ff} as the cyan ellipse.
This average gives $\chi^2/{\rm ndf} = 48/9$ ($P=\SN{3}{-7}$); the errors 
on $\lambda_+'$ and $\lambda_+''$ are scaled by 1.4 in this case, 
while the error on $\lambda_0$ is scaled by 2.2.
If instead of the NA48 $K_{e3}$-$K_{\mu3}$ average from \Tab{tab:Kl3ffdata}, 
only the $K_{e3}$ measurement from NA48 \cite{Lai:2004kb} is used, much greater 
consistency is observed. The resulting average is plotted as the yellow ellipse 
in \Fig{fig:Kl3ff}; this average gives $\chi^2/{\rm ndf} = 12.1/8$ 
($P=14.5\%$).

\begin{table*}
\center
\begin{tabular}{lcccc}
\hline\hline
Experiment & $\SN{\Lambda_+}{3}$ & $\lnC$ & $\rho$ & Ref. \\
\hline
KLOE   & $25.70\pm0.57$ & $0.2038\pm0.0241$ & $-0.26$ & \cite{Ambrosino:2007yz} \\
KTeV   & $25.09\pm0.44$ & $0.1915\pm0.0116$ & $-0.27$ & \cite{Abouzaid:2009ry} \\
NA48   & $24.60\pm1.47$ & $0.1354\pm0.0133$ & $-0.24$ & \cite{Lai:2007dx} \\
ISTRA+ & $26.13\pm0.52$ & $0.2084\pm0.0134$ & $-0.46$ & \cite{ISTRAprivate} \\
\hline
Average & $25.66\pm0.41$ & $0.2004\pm0.0091$ & $-0.33$ & \\
\hline\hline
\end{tabular}
\caption{Dispersive form factor parameters for $K_{\ell3}$ decays
from KLOE, KTeV, NA48, and ISTRA+.
For all experiments except NA48, the $K_{e3}$-$K_{\mu3}$ 
average is quoted by the experiment.
NA48 does not quote an average; the table lists our average of their results,
which has $S=2.0$ for $\Lambda_+$. 
The values from ISTRA+ have been converted for use with the notation 
of \Eqs{eq:Dispf} and~\andEq{eq:Dispfp}; the systematic uncertainties on the 
ISTRA+ measurements are derived from systematic studies using the pole fits.
Our average of the results from all four experiments, excluding the 
$K_{\mu3}$ data from NA48, is also listed.}
\label{tab:ffdis}
\end{table*}
\begin{figure}
\centering
\includegraphics[width=0.7\linewidth]{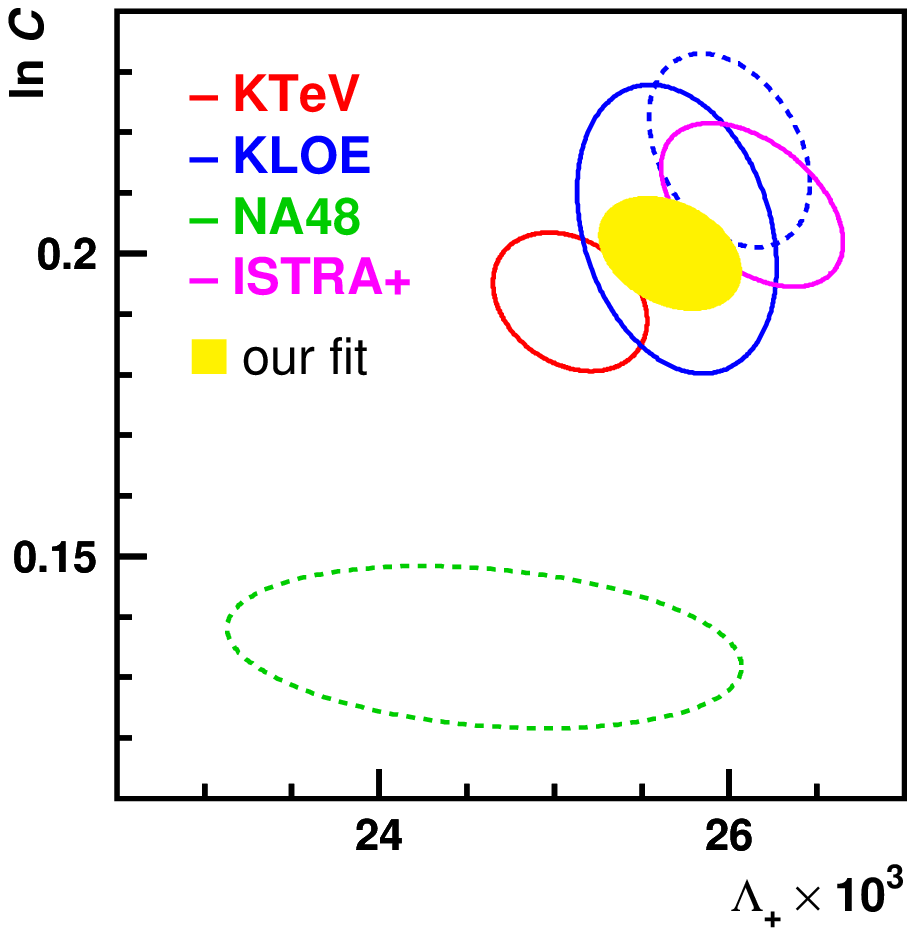}
\caption{
$1\sigma$ confidence contours for measurements of 
$\Lambda_+$ and \lnC\ from KLOE, KTeV, NA48, and ISTRA+.
For each experiment, the results for $K_{e3}$ and $K_{\mu3}$ decays are
averaged.
Our average of the results from all four experiments, excluding the
$K_{\mu3}$ data from NA48, is also shown. A new, preliminary KLOE result
\cite{Testa:2008xz} 
is shown as the dashed blue ellipse. This is not included in the fit.}
\label{fig:logc}
\end{figure}
\begin{sloppypar}
Apart from concerns about the consistency of the data set,
as described in \Sec{sec:disp}, the use of a 
linear parameterization for the scalar form factor is inherently problematic;
the use of a Class-I parameterization 
for the form factors such as 
the dispersive parameterization of \Eqs{eq:Dispf} and~\andEq{eq:Dispfp}
is greatly preferred.
Recently KLOE, KTeV, NA48, and ISTRA+ have obtained 
results for the parameters ($\Lambda_+$, \lnC) from fits to their 
$K_{\ell3}$ data.
These results are listed in \Tab{tab:ffdis} and shown graphically in 
\Fig{fig:logc}.
For each experiment, the results for $K_{e3}$ and $K_{\mu3}$ decays are
averaged.
As in \Tab{tab:Ke3pd}, in \Tab{tab:ffdis} the uncertainties on the values 
from the individual experiments do 
not include the contributions arising from the representation
of the form factor phase in the dispersive parameterization;
these contributions are common to all experiments and are included as
additional uncertainties of
\SN{0.30}{-3} on our average value of $\Lambda_+$ and
0.0040 on our average value of \lnC.

The use of the dispersive form factor parameterization clearly illustrates
the contrast between the $K_{\mu3}$ result from NA48 and those from the
other experiments.
The remaining measurements are in agreement.
Our average of the results from all four experiments,
excluding only the $K_{\mu3}$ data from NA48, is plotted as the yellow ellipse.
This average gives $\chi^2/{\rm ndf} = 5.6/5$ ($P=34.4\%$).
By contrast, if the $K_{\mu3}$ data from NA48 are included, the average 
gives $\chi^2/{\rm ndf} = 25.7/6$ ($P=0.026\%$).
On this basis, we exclude the NA48 $K_{\mu3}$ form factor results from the
averages used to calculate the phase space integrals for the evaluation of
\Vus\ and related tests. 
\end{sloppypar}

\begin{table}
\center
\begin{tabular}{lccc}
\hline\hline
Integral          & $\lambda_+',\lambda_+'',\lambda_0$ & $\Lambda_+, \lnC$ & Rel.\ diff. \\ \hline
$I(K^0_{e3})$       & 0.15457(20) & 0.15476(18) & $+0.12\%$ \\
$I(K^\pm_{e3})$     & 0.15894(21) & 0.15922(18) & $+0.18\%$ \\
$I(K^0_{\mu3})$     & 0.10266(20) & 0.10253(16) & $-0.13\%$ \\
$I(K^\pm_{\mu3})$   & 0.10564(20) & 0.10559(17) & $-0.05\%$ \\
$\rho(K_{e3},K_{\mu3})$ & $+0.56$ & $+0.38$ \\
\hline\hline
\end{tabular}
\caption{Comparison of phase space integrals evaluated from our averages of the results of quadratic-linear $(\lambda_+',\lambda_+'',\lambda_0)$ and dispersive 
$(\Lambda_+, \lnC)$ fits.}
\label{tab:integrals}
\end{table}
Table \ref{tab:integrals} lists the values of the phase space integrals
as computed from the results of our averages of the experimental form factor
data using the quadratic-linear and dispersive parameterizations.
(In the case of the dispersive parameterization, the uncertainties
arising from the representation of the form factor phase
are included in the overall uncertainty for each integral.)
For both parameterizations, the correlations between the uncertainties 
on the integrals are described by a matrix of the form
\begin{displaymath}
\left(
\begin{array}{cccc}
\ 1\ & \ 1\ & \ \rho\ & \ \rho\ \\
1 & 1 & \rho & \rho \\
\rho & \rho & 1 & 1 \\
\rho & \rho & 1 & 1
\end{array} 
\right).
\end{displaymath}
The table lists the values of the correlation coefficient $\rho$ for 
each parameterization.

As seen in the table, when evaluated from the dispersive fit 
results, the integrals tend to be very slightly different than they are 
when evaluated from the quadratic-linear fit results.
Nowhere is this difference greater than 0.2\%.
As expected, the $K_{\mu3}$ integrals are slightly smaller
when obtained from the dispersive fit results (\Sec{sec:FFpar}).
Given the advantages of the dispersive
parameterization, we use the integrals calculated from the dispersive fit 
results for the evaluation of \Vus\ and related tests.

\subsubsection{The $K_{\mu3}$ scalar form factor and tests of chiral 
perturbation theory}
\label{sec:CTtest}

\begin{figure}
\centering
\includegraphics[width=0.75\linewidth]{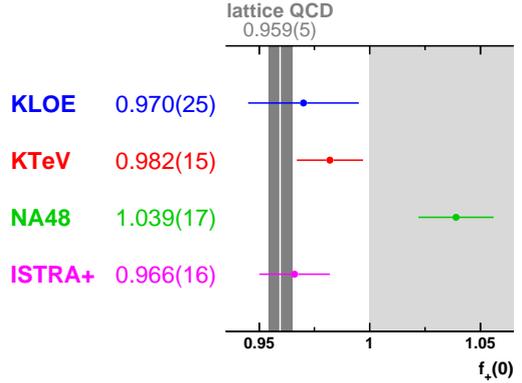}
\caption{Values for \fp\ from measurements of \lnC\ from different
experiments, using the Callan-Treiman relation and  $\FKpi=1.193(6)$. 
The lattice QCD result $\fp=0.959(5)$ is also shown.}
\label{fig:CTtest}
\end{figure}
\begin{sloppypar}
Given a value for \FKpi, the Callan-Treiman relation (\Eq{eq:CTrel}) 
can be used to obtain a value for \fp\ from a measurement of \lnC,
providing a test of consistency between scalar form factor measurements and 
lattice calculations.
\Figure{fig:CTtest} shows the values for \fp\
corresponding to the measurements of \lnC\ from different experiments,
using the CT relation and $\FKpi = 1.193(6)$ (\Sec{sec:FKpi}).
For this exercise,
the uncertainties on each value of \lnC\ include the common contribution
from the parameterization of the form factor phase.
The lattice QCD value, $\fp = 0.959(5)$ (\Eq{eq:fpQCD}),
is also shown.
The measurements of \lnC\ from KLOE, KTeV, and ISTRA+ give
values for \fp\ that are essentially consistent with the lattice 
estimate, although they tend to be a little larger. The NA48 result,
on the other hand, gives 
$\fp = 1.039(17)$, which is in contrast with the theoretical expectation
of Fubini and Furlan that $\fp < 1$ \cite{Fubini:1970ss}.
As noted in \Sec{sec:Kl3ff}, we exclude the NA48 $K_{\mu3}$ form factor 
measurements from the averages used to calculate the phase space integrals 
for the evaluation of \Vus\ and related tests.
Our resulting world-average values for \Lp\ and \lnC\ are listed in 
\Tab{tab:ffdis}.
The value $\lnC = 0.2004(91)$, when used with $\FKpi = 1.193(6)$,  
gives $\fp = 0.974(12)$. More generally, the experimental data on \lnC\ 
alone give $(\FKpi)/\fp=1.225(14)$. This result is completely 
independent of any information from lattice estimates.
\end{sloppypar}

\begin{table}
\centering
\begin{tabular}{lc|rrr}
\hline\hline
Parameter & Value & \multicolumn{3}{c}{Corr.\ matrix (\%)} \\
\hline
$\SN{\Lp}{3}$ & $25.7\pm0.5$ & $-27$ & $+10$ & $-10$ \\
$\lnC$        & 0.208(8)     &       & $-38$ & $+38$ \\
$\fp$         & 0.961(5)     &       &       & $+19$ \\
$\FKpi$       & 1.191(6)     &       &       &       \\
\hline\hline
\end{tabular}
\caption{Results of a fit to the experimental average value for \lnC\ 
(including a preliminary KLOE measurement) and lattice results for \fp\ 
and \FKpi, using the constraint provided by the Callan-Treiman theorem.}
\label{tab:ffsct}
\end{table}
Alternatively, one can perform a fit to the world-average value of \lnC,
together with the lattice determinations $\FKpi = 1.193(6)$ and 
$\fp = 0.959(5)$, using the constraint given by the CT
relation. When performing such a fit, we make use of a recent preliminary 
measurement from KLOE \cite{Testa:2008xz} 
of the dispersive $K_{\mu3}$ form factor parameters.
This measurement, which is illustrated as the dashed ellipse in 
\Fig{fig:logc}, is used here in place of the published KLOE 
measurement in \Tab{tab:ffdis}. Although preliminary, we make use of
this measurement here as we are only interested in demonstrating the 
power of the consistency test offered by the CT relation.
The average of the dispersive $K_{\mu3}$ form factor parameters using
the new KLOE measurement gives 
$\Lp = \SN{(25.78\pm0.40)}{-3}$ and
$\lnC = 0.2034(86)$, with $\rho(\Lp,\lnC) = -0.34$ and 
$\chi^2/{\rm ndf} = 7.4/5$ ($P = 19.2\%$).
The value of \lnC\ is then fitted together with the lattice inputs as 
described above to obtain the results in \Tab{tab:ffsct}.
The fit gives $\chi^2/{\rm ndf} = 0.78/1$ ($P = 38\%$), confirming the 
agreement between the experimental measurements of \lnC\ and
the lattice determinations of \fp\ and \FKpi.
The values of \fp\ and \FKpi\ move only slightly, and the 
uncertainties on these values are slightly decreased.

\section{Physics results}
\label{sec:res}

\subsection{\mathversion{bold}Determination of \Vusf}
\label{sec:vusf}

\begin{table*}
\centering
\begin{tabular}{lcccccc|rrrr}
\hline\hline
Mode                & \Vusf\       & \% err & BR   & $\tau$ & $\Delta$ & Int & \multicolumn{4}{c}{Correlation matrix (\%)} \\
\hline
$K_L \to \pi e \nu$     & 0.2163(6)    & 0.26   & 0.09 & 0.20  & 0.11  & 0.06 & $+55$ & $+10$ & $+3$ & $0$   \\
$K_L \to \pi \mu \nu$   & 0.2166(6)    & 0.29   & 0.15 & 0.18  & 0.11  & 0.08 &       & $+6$  & $0$  & $+4$  \\
$K_S \to \pi e \nu$     & 0.2155(13)   & 0.61   & 0.60 & 0.03  & 0.11  & 0.06 &       &       & $+1$ & $0$   \\
$K^\pm \to \pi e \nu$   & 0.2160(11)   & 0.52   & 0.31 & 0.09  & 0.40  & 0.06 &       &       &      & $+73$ \\
$K^\pm \to \pi \mu \nu$ & 0.2158(14)   & 0.63   & 0.47 & 0.08  & 0.39  & 0.08 &       &       &      &       \\
Average                & 0.2163(5)  &    &  &   &   & \\
\hline\hline 
\end{tabular}
\caption{Values of \Vusf\ as determined from each kaon decay mode, 
with approximate contributions to relative uncertainty (\% err) from
branching ratios (BR), lifetimes ($\tau$), combined effect of 
\dEM{K\ell} and \dSU{K\ell} ($\Delta$), and phase space integrals (Int).}
\label{tab:Vusf0}
\end{table*}
\begin{figure}
\centering
\includegraphics[width=0.6\linewidth]{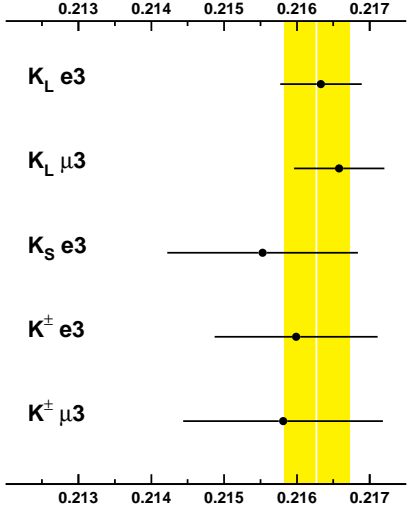}
\caption{Comparison of values for \Vusf\ for all channels. 
Our average is indicated by the yellow band.} 
\label{fig:Vusf0}
\end{figure}
For each of the five decay modes for which rate measurements exist, we use 
\Eq{eq:Mkl3} to evaluate \Vusf\ from the decay rate data in
\Tabs{tab:KLBR}, \ref{tab:KSfit}, and \ref{tab:KpmBR}, the phase space 
integrals from dispersive fits in \Tab{tab:integrals}, 
the long-distance radiative corrections 
in \Tab{tab:Kl3radcorr}, and the \SU{2}-breaking corrections of 
\cite{Kastner:2008ch} discussed in \Sec{sec:isobreak}. We keep track of the 
correlations between the uncertainties on the values of \Vusf\ from different 
modes arising from the use of common corrections and from correlations in 
the input data set (e.g., from the outputs of the fits to BR and lifetime
measurements).
The resulting values of \Vusf\ are listed in \Tab{tab:Vusf0}
and illustrated in \Fig{fig:Vusf0}.
The principal experimental result of this review is the average value   
\begin{equation}
\Vusf=0.2163(5),
\label{eq:vusfres}
\end{equation}
which has an uncertainty of about of $0.2\%$.  
The results from the five modes are in good agreement; the
fit gives $\chi^2/{\rm ndf}=0.77/4$ ($P=94\%$).
\Table{tab:Vusf0} gives an approximate breakdown of the sources
contributing to the total uncertainty on the determination of 
\Vusf\ from each mode. 
The best single determinations of \Vusf\ are from the $K_{e3}$ and 
$K_{\mu3}$ modes, with $K_{e3}$ giving the slightly better determination
since there is no contribution from uncertainties on the parameters of
the scalar form factor. The limited precision of the value for \Vusf\ from 
$K_{e3}$ decays of the $K_S$ decays is entirely determined by the
experimental uncertainty on the corresponding BR, which is dominantly 
statistical. A better measurement of the BR for this decay would allow
knowledge of \Vusf\ to be significantly improved, since $\tau_{K_S}$ is
known very precisely. The values of \Vusf\ from charged kaon decays
are currently limited in precision both by experimental uncertainties 
in the corresponding BRs and by the uncertainty in the 
theoretical estimate of \dSU{K^+\pi^0}.

\subsection{Accuracy of isospin-breaking corrections}
\label{sec:ib}

The average values for \Vusf\ can be computed separately for charged and 
neutral kaon decays. The ratio of these values (calculated without 
applying \SU{2}-breaking corrections) then gives an experimental 
estimate of \dSU{K^+\pi^0} for comparison with the
estimate from chiral perturbation theory.
We obtain
\begin{equation}
\delta_{\SU{2},\:{\rm exp}}^{K^+\pi^0} = 0.027 \pm 0.004,
\label{eq:IBexp}
\end{equation}
in good agreement with the theoretical estimate of \Ref{Kastner:2008ch},
$\dSU{K^+\pi^0} = 0.029 \pm 0.004$ (\Eq{eq:isobrk}). We observe that 
the uncertainty on the theoretical value of $\dSU{K^+\pi^0}$
contributes significantly to the overall uncertainties on the values
of \Vusf\ for charged kaon decay modes (see \Tab{tab:Vusf0}).

\subsection{Lepton universality}
\label{sec:LU}

Comparison of the values of \Vusf\ computed separately for $K_{e3}$ 
and $K_{\mu3}$ modes provides a test of lepton universality. 
Specifically, 
\begin{equation}
r_{\mu e}
= \frac{[\Vusf]^2_{\mu3,\:{\rm exp}}}
       {[\Vusf]^2_{e3,\:{\rm exp}}} 
= \frac{\Gamma_{K\mu3}}{\Gamma_{Ke3}}
  \frac{I_{e3}\,(1 + 2\dEM{Ke})}{I_{\mu3}\,(1 + 2\dEM{K\mu})}.
\label{eq:rme}
\end{equation}
(Here, $\dEM{Ke}$ and $\dEM{K\mu}$ are the
electromagnetic corrections for kaons of the charge state under consideration.)
By comparison with \Eq{eq:Mkl3}, $r_{\mu e}$ is equal to the ratio
$g_\mu^2/g_e^2$, with $g_\ell$ the coupling strength at the 
$W\to\ell\nu$ vertex. In the SM, $r_{\mu e} = 1$.

Before the advent of the new BR measurements described in 
\Secs{sec:KL} and~\ref{sec:Kpm}, the values of \Vusf\ from 
$K_{e3}$ and $K_{\mu3}$ rates were in substantial disagreement. 
Using the $K_L$ and $K^\pm$ BRs from the 2004 edition of the PDG 
compilation \cite{Eidelman:2004wy} 
(and assuming current values for the $I_{\ell3}$ and 
$\dEM{K\ell}$), we obtain $r_{\mu e} = 1.013(12)$ for 
$K^\pm$ decays and 1.040(13) for $K_L$ decays.

As noted in \Sec{sec:KL}, the new BR measurements procure much better
agreement. From the entries in \Tab{tab:Vusf0}, we calculate $r_{\mu e}$ 
separately for charged and neutral modes (including the value of \Vusf\ from
$K_S\to\pi e\nu$ decays, though this has little impact) 
and obtain 0.998(9) and 1.003(5), respectively.
The results are compatible; the average value is $r_{\mu e} = 1.002(5)$.
As a statement on the lepton-flavor universality hypothesis,
we note that the sensitivity of this test approaches that
obtained with $\pi\to\ell\nu$ decays
($(r_{\mu e}) = 1.0042(33)$ \cite{RamseyMusolf:2007yb})
and $\tau\to\ell\nu\bar{\nu}$ decays 
($(r_{\mu e}) = 1.000(4)$ \cite{Davier:2005xq}).
Alternatively, if the lepton-universality hypothesis is assumed to be
true, the equivalence of the values of \Vusf\ from $K_{e3}$ and $K_{\mu3}$
demonstrates that the calculation of the long-distance corrections
$\dEM{K\ell}$ is accurate to the per-mil level.

\subsection{\mathversion{bold}Determination of $\Vusd\times\FKpi$}
\label{sec:fKpVusd}

As noted in \Sec{sec:Pl2}, \Eq{eq:Mkl2} 
allows the ratio $\Vusd\times\FKpi$ to be determined from experimental
information on the radiation-inclusive $K_{\ell2}$ and $\pi_{\ell2}$ decay 
rates. The limiting uncertainty is that from \BR{K_{\mu2(\gamma)}}, which is 
0.28\% as per \Tab{tab:KpmBR}.
Using this, together with the value of $\tau_{K^\pm}$ from the same fit and 
$\Gamma(\pi^\pm\to\mu^\pm\nu)=38.408(7)~\mu{\rm s}^{-1}$ \cite{Amsler:2008zzb}
we obtain
\begin{equation}
\Vusd \times \FKpi = 0.2758(5).
\label{eq:vusvudres}
\end{equation}

\subsection{Test of CKM unitarity}
\label{sec:univ}

\begin{figure}
\centering
\includegraphics[width=0.8\linewidth]{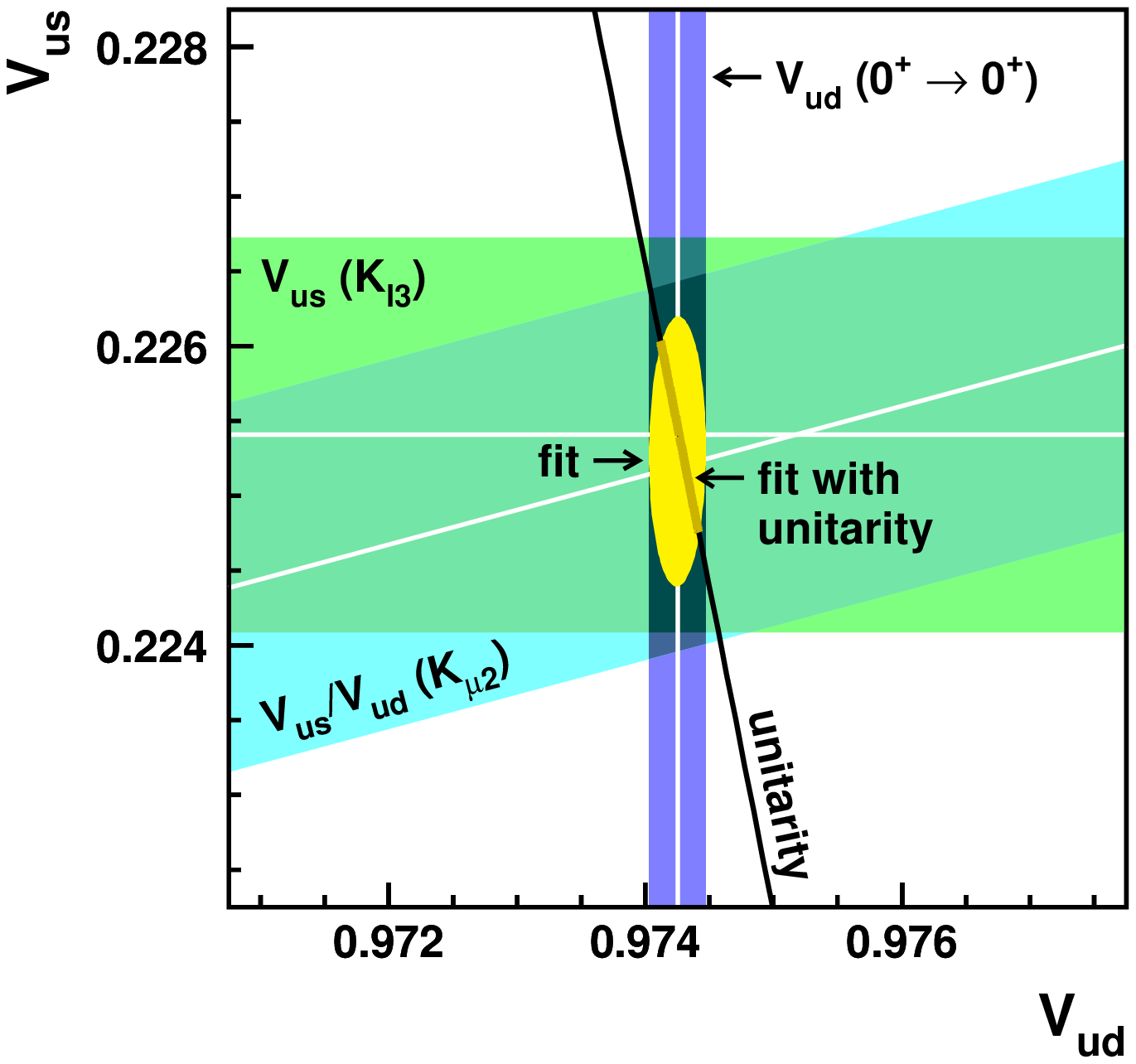}
\caption{Results of fits to \Vud, \Vus, and \Vusd.}
\label{fig:vusuni}
\end{figure}
\begin{sloppypar}
We determine \Vus\ and \Vud\ from a fit to the results obtained above.
As starting points, we use the value $\Vusf = 0.2163(5)$ given 
in \Tab{tab:Vusf0}, together with the lattice QCD estimate 
$\fp = 0.959(5)$ (\Eq{eq:fpQCD}).
We also use the result $\Vusd \times \FKpi = 0.2758(5)$ discussed in
\Sec{sec:fKpVusd} together with the lattice estimate
$\FKpi = 1.193(6)$ (\Sec{sec:FKpi}).
Thus we have
\begin{align}
\begin{split}
\Vus  &= 0.2254(13) \qquad [K_{\ell 3}~{\rm only}], \\
\Vusd &= 0.2312(13) \qquad [K_{\ell 2}~{\rm only}].
\end{split}
\label{eq:VusVudIn}
\end{align}
Finally, we use the evaluation 
$\Vud = 0.97425(22)$ from a recent survey \cite{Hardy:2008gy} of
half-life, decay-energy, and BR measurements related to 20 superallowed 
$0^+ \to 0^+$ nuclear beta decays, which
includes a number of new, high-precision Penning-trap measurements of 
decay energies, as well as the use of recently improved electroweak 
radiative corrections \cite{Marciano:2005ec} and new isospin-breaking
corrections \cite{Towner:2007np}, in addition to other improvements over 
past surveys by the same authors.
Our fit to these inputs gives 
\begin{align}
\begin{split}
\Vud &= 0.97425(22), \\
\Vus &= 0.2253(9) \qquad [K_{\ell3}, K_{\ell2}, 0^+\to0^+],
\label{eq:VusVud}
\end{split}
\end{align}
with $\chi^2/{\rm ndf} = 0.014/1$ ($P=91\%$) and negligible correlation between
\Vud\ and \Vus. 
With the current world-average value, $\Vub = 0.00393(36)$ 
\cite{Amsler:2008zzb}, the first-row unitarity sum is then 
$ \Delta_{\rm CKM} = \Vud^2 + \Vus^2 + \Vub^2 - 1 = -0.0001(6)$;
the result is in striking agreement with the unitarity hypothesis.
(Note that the contribution to the sum from \Vub\ is 
essentially negligible.)
As an alternate expression of this agreement, we may state a value for
$G_{\rm CKM} = G_\mu \sqrt{\Vud^2+\Vus^2+\Vub^2}$. We obtain
\begin{equation}
G_{\rm CKM} = 1.16633(35) \times 10^{-5}~{\rm GeV}^{-2},
\label{eq:GCKM}
\end{equation}
with $G_\mu = 1.166371(6) \times 10^{-5}~{\rm GeV}^{-2}$ \cite{Chitwood:2007pa}.
\end{sloppypar}

It is also possible to perform the fit with the unitarity constraint
included, increasing by one the number of degrees of freedom. The constrained
fit gives 
\begin{equation}
\Vus=\sin\,\theta_C=\lambda=0.2254(6) \qquad [{\rm with~unitarity}]
\end{equation}
and $\chi^2/{\rm ndf}=0.024/2$ ($P=99\%$). 
This result and that obtained above without assuming unitarity are both 
illustrated in \Fig{fig:vusuni}.

At this point, using \Eq{eq:dckmnp} and the phenomenological value 
$ \Delta_{\rm CKM}  = -0.0001(6)$,  it is possible to set bounds on the 
effective scale of the four operators that parameterize new physics
contributions to $\Delta_{\rm CKM}$. We obtain
\begin{equation}
\Lambda > 11~{\rm TeV} \qquad (90\%~{\rm C.L.}). 
\label{eq:LB1}
\end{equation}
As noted in \Ref{Cirigliano:2009wk}, 
for the operators $O_{ll}^{(3)}, O_{\varphi l}^{(3)}$, and $O_{\varphi q}^{(3)}$  
(see \Eqs{eq:operators}), this constraint is at the same level as the
constraints from $Z$-pole measurements. 
For the four-fermion operator $O_{lq}^{(3)}$,
$\Delta_{\rm CKM}$ improves upon existing bounds from LEP2 by an order 
of magnitude.

\subsection{Bounds on non-helicity-suppressed amplitudes}
\label{sec:bounds}

\begin{figure}
\centering
\includegraphics[width=0.8\linewidth]{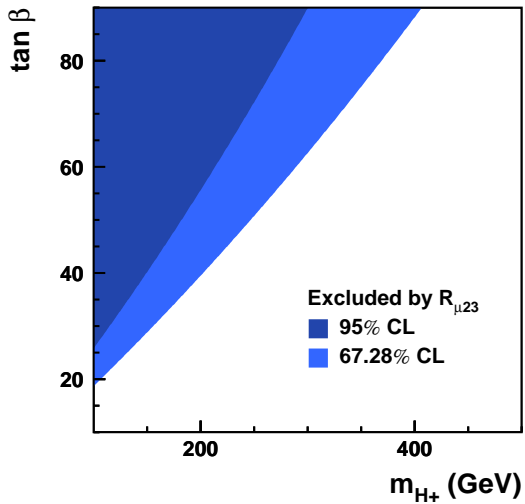}
\caption{Regions in the $(m_{H^\pm}, \tan \beta)$ plane 
in two-Higgs-doublet models excluded by the present 
result for $R_{\mu23}$.}
\label{fig:2HDM}
\end{figure}
As noted in \Sec{sec:2HDM}, an empirical value for the ratio $R_{\mu23}$
(\Eq{eq:Rl23}) can be used to exclude regions of the $(m_{H^\pm}, \tan \beta)$
parameter space in models with two Higgs doublets, such as the 
minimal supersymmetry extension of the SM
(\Eq{eq:2HDM}). Operatively, we evaluate $R_{\mu23}$ via a fit akin to that 
described in \Sec{sec:univ}, but with separate parameters accounting for 
the values of \Vus\ from $K_{\ell3}$ and $K_{\mu2}$ decays. 
The fit then has three free parameters:
the value of \Vus\ from $K_{\ell3}$ decays, the value of \Vusd\ from 
$K_{\mu2}$ decays, and the value of \Vud\ from $0^+\to0^+$ nuclear beta decays.
The input values used for \Vus\ and \Vusd\ are from \Eq{eq:VusVudIn}
and include the relevant lattice constants. The contribution to
non-helicity-suppressed $K_{\ell3}$ decays from charged Higgs exchange 
is negligible, so we include as a constraint
in the fit the first-row unitarity condition on the value of \Vus\ from
$K_{\ell3}$ decays: $\Vud^2 + \Vus^2_{K_{\ell3}} + \Vub^2= 1$. 
Expressing the results of the fit in terms of \Vus\ from $K_{\ell3}$ decays and
the ratio $R_{\mu23}$, we obtain
\begin{align}
\begin{split}
\Vus &= 0.2254(8) \qquad 
[K_{\ell3}, 0^+\to0^+, {\rm unitarity}],\\
R_{\mu23} &= 0.999(7) \qquad [K_{\mu2}].
\end{split}
\end{align}
The fit gives $\chi^2/{\rm ndf}=0.0003/1$ ($P=99\%$), 
with $\rho=-0.55$ between the parameter uncertainties in the stated basis. 
The regions of the $(m_{H^\pm}, \tan \beta)$ parameter space 
in models with two Higgs doublets excluded at the $1\sigma$ and 95\% CLs 
by this result for $R_{\mu23}$ are shown as the shaded area in \Fig{fig:2HDM}.
The bound is obtained setting $\epsilon_0=1/16\pi^2$ in \Eq{eq:2HDM}, 
as expected in the MSSM. Note that this result excludes the region 
at low $m_{H^\pm}$  and large  $\tan \beta$ favoured by 
$B\to\tau\nu$ \cite{Bona:2009cj}.

\subsection{\mathversion{bold}Determination of Standard Model values for \fp\ and \FKpi\ 
from experimental data}

\begin{figure}
\centering
\includegraphics[width=0.8\linewidth]{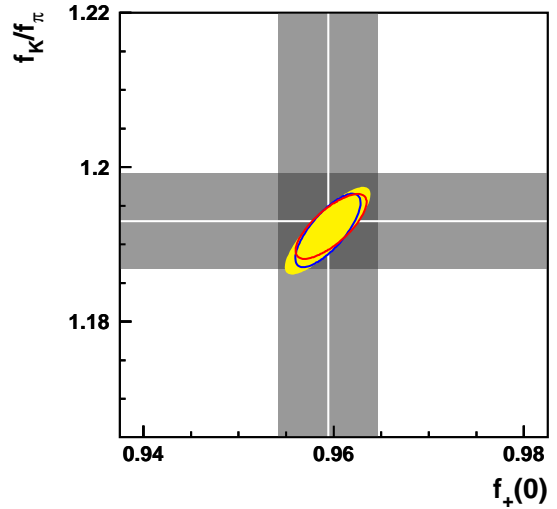}
\caption{Results of fit for \fp\ and \FKpi, given experimental data and first-row CKM unitarity: (yellow) no input from lattice; (blue) $\fp = 0.959(5)$ as input to the fit and no input for \FKpi; (red) $\FKpi = 1.193(6)$ as input and no input for \fp. The grey bands illustrate the reference values for \fp\ and \FKpi.}
\label{fig:const}
\end{figure}
\Equation{eq:Mkl2}, which in the SM relates the ratio of 
$K_{\ell2}$ and $\pi_{\ell2}$ decay rates to the ratio $\Vusd\times\FKpi$,
can be rewritten
\begin{equation}
Q_{\ell2} =
\frac{(\Vusf)^2}{\Vud^2} \times
\frac{1}{\fp^2} \times
\frac{\FK^2}{\Fpi^2},
\label{eq:Ql2}
\end{equation}
where $Q_{\ell2}$ is the ratio of $\Gamma(K_{\ell2(\gamma)})$ to 
$\Gamma(\pi_{\ell2(\gamma)})$ with phase space differences factored out and
radiative corrections applied.
As per the discussion in \Secs{sec:Pl2} and~\ref{sec:fKpVusd},
for $K_{\mu2(\gamma)}$ and $\pi_{\mu2(\gamma)}$ decays, $Q_{\mu2} = 0.07604(26)$.

\Equation{eq:Ql2} can be used together with the experimentally derived
values for $Q_{\mu2}$ from $K_{\mu2}$ and $\pi_{\mu2}$ rates, \Vusf\ from 
$K_{\ell3}$ decays, and $\Vud$ from $0^+ \to 0^+$ nuclear beta decays to 
determine the values of the hadronic constants \fp\ and \FKpi.
A straightforward calculation gives
\begin{equation}
(\FKpi)/\fp = 1.242(4).
\end{equation}
This value essentially depends only on experimental $K_{\mu2}$ and 
$\pi_{\mu2}$ rate data and radiative corrections. CKM unitarity is not assumed,
although the equality of \Vus\ in $K_{\mu2}$ and $K_{\ell3}$ decays is.
If we further assume the validity of our reference lattice value 
$\fp = 0.959(5)$, we obtain $\FKpi = 1.192(8)$, while if we assume
the validity of our reference value 
$\FKpi = 1.193(6)$, we obtain $\fp = 0.960(6)$.

To go further, we impose first-row CKM unitarity as an additional constraint.
We can then perform a fit with $Q_{\mu2}$, \Vusf, \Vud, \fp, and \FKpi\ as
parameters, and two constraints: $\Vud^2 + \Vus^2 +\Vub^2 = 1$ and \Eq{eq:Ql2}.
Up to two of the parameters can be left free for a solution with no degrees
of freedom. Taking these to be \fp\ and \FKpi, we obtain
\begin{align}
\begin{split}
\fp   &= 0.959(5), \\
\FKpi &= 1.192(6) \qquad [{\rm with~unitarity}],
\end{split}
\end{align}
with a correlation coefficient of $\rho = +0.84$. This solution 
is illustrated as the yellow ellipse in \Fig{fig:const}. Once again, we note
that this result is obtained without any recourse to prior estimates of 
\fp\ or \FKpi. The lattice values of these constants are illustrated as 
the grey bands in \Fig{fig:const} for the purposes of comparison only.

With either of the reference values of \fp\ or \FKpi\ as an additional 
input, the fit has one degree of freedom. 
With $\fp = 0.959(5)$ as input, the fit gives 
$\Vud = 0.97425(18)$, $\Vusf = 0.2163(4)$, $\fp = 0.9594(34)$, and 
$\FKpi = 1.192(5)$, with $\rho=+0.76$ the correlation coefficient between
\fp\ and \FKpi, and $\chi^2 = \SN{3}{-4}/1$ ($P=99\%$). This is illustrated as 
the blue ellipse in \Fig{fig:const}.
With $\FKpi = 1.193(6)$ as input instead, the fit gives 
$\Vud = 0.97427(17)$, $\Vusf = 0.2163(5)$, $\fp = 0.960(4)$, and 
$\FKpi = 1.192(4)$, with $\rho=+0.75$ and $\chi^2 = 0.023/1$ $(P=88\%)$. 
This is illustrated as the red ellipse in \Fig{fig:const}. In summary,
our chosen reference values for \fp\ and \FKpi\ are a near-perfect match
with experimental data and SM assumptions.

\section{Conclusions}

At present, the experimental precision of data on leptonic and 
semileptonic kaon decays is nicely matched below the percent level 
by the theoretical precision,
allowing us to perform very precise measurements
of SM parameters and to set stringent bounds on 
beyond-SM physics. In this work, we have presented an 
updated analysis of the following three main subjects:  
(i) the overall determination of \Vus, with and without
imposing CKM unitarity; 
(ii) the comparison between the values of \Vus\
as determined from $K_{\ell3}$ and  $K_{\mu2}$ decay data, and the corresponding 
constraints on deviations from the $V-A$ structure;
(iii) tests of lepton universality in $K_{\ell3}$ 
decays. The final results of our analysis are 
presented in detail in \Sec{sec:res} and will not be repeated here. 
As a general conclusion, we simply emphasize that the \order{10~{\rm TeV}}
bound on the scale of new physics (\Eq{eq:LB1}), which
follows from the verification of the first-row CKM unitarity condition, 
represents one of the most stringent constraints on
physics beyond the Standard Model.

\begin{sloppypar}
In addition to the three main topics mentioned above, our analysis has 
allowed some interesting cross-checks of 
both analytic and lattice-based theoretical results relevant 
to kaon physics, most notably the determination of the 
mass-induced isospin-breaking effect between charged 
and neutral semileptonic kaon decays (\Eq{eq:IBexp}) 
and the indirect determination of the kaon
form factors shown in \Fig{fig:const}. We emphasize
that both of these results, which agree well with the current 
theoretical estimates, are strictly valid only under a 
pure SM hypothesis. 
\end{sloppypar}

\section*{Acknowledgments}

\begin{sloppypar}
The authors thank P.~Franzini and all other members of the 
FlaviaNet Kaon Working Group, as well as 
V.~Lubicz and A.~J\"uttner of the FlaviaNet Lattice Averaging Group (FLAG),
for useful discussion, comments, and suggestions.
This work is supported by the European Union Sixth Framework Programme
under contract MTRN-CT-2006-035482, FlaviaNet.
E.P. acknowledges financial support from MEC (Spain) under grant
FPA2007-60323. F.M. acknowledges financial support from projects
FPA2007-66665, 2009SGR502.
E.P. and F.M. acknowledge support from
the Spanish Consolider-Ingenio 2010 Programme CPAN
(CSD2007-00042). G.I. acknowledges support from the
Technische Universit\"at M\"unchen - Institute for Advanced Study, 
funded by the German Excellence Initiative. 
\end{sloppypar}

\begin{appendix}

\section{\mathversion{bold}Polynomial approximations for the phase space integrals $I_{K\ell}$}
\label{sec:polIk}

In this appendix, we give the explicit forms used to evaluate the 
phase space integrals $I_{K\ell}$ using the form factor parameters obtained
from fits with different parameterizations. In all cases, we insert the 
expressions for $\bar{f}_{+,0}(t)$ into the integrand of \Eq{eq:IK} and obtain
a polynomial expression in the form factor parameters for the integral 
$I_{K\ell}$, with coefficients depending on the mass values $m_K$ and $m_\ell$. 

The form of the phase space integral used with the polynomial parameterization 
up to cubic terms in $t$ is
\begin{align}
\begin{split}
I_{K\ell} &=
 z_0 + z_1\lp' + z_2(\lp'^2+\lp'') + z_3\lp'\lp'' + z_4\lp''^2 \\ 
 & + z_5\lp''' + z_6\lp'\lp''' + z_7\lp''\lp''' + z_8\lp'''^2 \\
 & + z_9\lz' + z_{10}(\lz'^2+\lz'') + z_{11}\lz'\lz'' + z_{12}\lz''^2 \\
 & + z_{13}\lz''' + z_{14}\lz'\lz''' + z_{15}\lz''\lz''' + z_{16}\lz'''^2.
\label{eq:Ipoly}
\end{split}
\end{align}
The numerical values of the coefficients $z$ are listed in \Tab{tab:z}.
For $K_{e3}$ decays, we take $z_9\ldots z_{17}$ to be zero.
We also use the above 
form with to evaluate phase space integrals from the results of fits using
the pole parameterization. In this case, we first perform a Taylor
expansion of \Eq{eq:pole} up to cubic terms.
\begin{table*}
\center
\begin{tabular}{lcccc|lcc}
\hline\hline
Coeff. & $I(K^\pm_{e3})$ & $I(K^0_{e3})$ & $I(K^\pm_{\mu3})$ & $I(K^0_{\mu3})$ &
Coeff. & $I(K^\pm_{\mu3})$ & $I(K^0_{\mu3})$ \\
\hline
$z_0$ & 0.14478436 & 0.14081706 & 0.09345981 & 0.09083246 & $z_9$    & 0.21195460 & 0.20545016 \\
$z_1$ & 0.50059594 & 0.48654454 & 0.30229496 & 0.29365924 & $z_{10}$ & 0.37262925 & 0.36033462 \\
$z_2$ & 0.69266614 & 0.67217878 & 0.48877555 & 0.47401926 & $z_{11}$ & 1.52866642 & 1.47403632 \\
$z_3$ & 2.39194070 & 2.31634083 & 1.80851699 & 1.75020047 & $z_{12}$ & 1.73806537 & 1.67074681 \\
$z_4$ & 2.35425967 & 2.27427126 & 1.84604117 & 1.78213612 & $z_{13}$ & 0.50955547 & 0.49134544 \\
$z_5$ & 0.79731357 & 0.77211361 & 0.60283900 & 0.58340016 & $z_{14}$ & 2.31742049 & 2.22766241 \\
$z_6$ & 3.13901290 & 3.03236168 & 2.46138823 & 2.37618149 & $z_{15}$ & 5.65071063 & 5.41411258 \\
$z_7$ & 6.74124302 & 6.49464837 & 5.40291925 & 5.20184325 & $z_{16}$ & 4.82367980 & 4.60616108 \\
$z_8$ & 5.13492355 & 4.93286555 & 4.17536001 & 4.00843393 &        &            &           \\
\hline\hline
\end{tabular}
\caption{Coefficients for the evaluation of phase space integrals from 
polynomial form factor parameters, for use with \Eq{eq:Ipoly}.}
\label{tab:z}
\end{table*}

The form of the phase space integral used with the dispersive 
parameterizations of \Eqs{eq:Dispf} and~\andEq{eq:Dispfp} is
\begin{align}
\begin{split}
I_{K\ell} &= c_0 + c_1\Lp + c_2\Lp^2 + c_3 \Lp^3 + c_4 \Lp^4 \\
        & + c_5\lnC + c_6(\lnC)^2 + c_7(\lnC)^3 + c_8(\lnC)^4.
\label{eq:Idisp}
\end{split}
\end{align}
The numerical values of the coefficients $c$ are listed in \Tab{tab:c}.
Again, for $K_{e3}$ decays, we take $c_5\ldots c_8$ to be zero.
\begin{table*}
\center
\begin{tabular}{lcccc|lcc}
\hline\hline
Coeff. & $I(K^\pm_{e3})$ & $I(K^0_{e3})$ & $I(K^\pm_{\mu3})$ & $I(K^0_{\mu3})$ &
Coeff. & $I(K^\pm_{\mu3})$ & $I(K^0_{\mu3})$ \\
\hline
$c_0$ & 0.14530227 & 0.14125098 & 0.09323787 & 0.09059706 & $c_5$ & 0.01797945 & 0.01724105 \\
$c_1$ & 0.50409859 & 0.48947133 & 0.30495081 & 0.29587728 & $c_6$ & 0.00545007 & 0.00515380 \\
$c_2$ & 1.39922105 & 1.35593685 & 0.98846227 & 0.95712982 & $c_7$ & 0.00128587 & 0.00119853 \\
$c_3$ & 3.22924652 & 3.12170948 & 2.44345185 & 2.36027254 & $c_8$ & 0.00025237 & 0.00023179 \\
$c_4$ & 6.36982500 & 6.14083977 & 6.53169253 & 4.81427789 &       &            & \\
\hline\hline
\end{tabular}
\caption{Coefficients for the evaluation of phase space integrals from
dispersive form factor parameters, for use with \Eq{eq:Idisp}.}
\label{tab:c}
\end{table*}

\end{appendix}

\bibliographystyle{flavia}
\bibliography{flavia09}

\end{document}